\label{key}
\documentclass[a4paper,fleqn,usenatbib]{mnras}

\usepackage{newtxtext,newtxmath}

\usepackage[T1]{fontenc}
\usepackage{ae,aecompl}
\usepackage{xspace}

\usepackage{graphicx}	
\graphicspath{{figures/}}
\usepackage{siunitx}
\DeclareSIUnit\gauss{G}
\DeclareSIUnit\erg{erg}
\DeclareSIUnit\parsec{pc}
\usepackage{amsmath}	
\usepackage{physics}
\usepackage[normalem]{ulem}
\usepackage{xparse}

\def\Fermi{{\em Fermi}\xspace}
\def\del#1{{}}

\title[Cosmic ray electrons in SN~1006]{Evolution and observational signatures of the cosmic ray electron spectrum in SN~1006}

\author[G. Winner et al.]
{Georg Winner,$^{1, 2}$\thanks{E-mail: gwinner@aip.de (GW), cpfrommer@aip.de (CP)}
Christoph Pfrommer,$^{1}$
Philipp Girichidis,$^{1}$
Maria Werhahn,$^{1}$
\newauthor
Matteo Pais$^{1, 2}$
\\
$^{1}$Leibniz-Institut f{\"u}r Astrophysik Potsdam (AIP), An der Sternwarte 16, 14482 Potsdam, Germany\\
$^{2}$Fakult{\"a}t f{\"u}r Physik und Astronomie, Universit{\"a}t Heidelberg, Im Neuenheimer Feld 226, 69120  Heidelberg, Germany\\
}

\date{Accepted 2020 September 22. Received 2020 September 14; in original form 2020 June 9}

\pubyear{2020}

\begin{document}
\label{firstpage}
\pagerange{\pageref{firstpage}--\pageref{lastpage}}
\maketitle

\begin{abstract}
Supernova remnants (SNRs) are believed to be the source of Galactic cosmic rays
(CRs). SNR shocks accelerate CR protons and electrons which reveal key insights
into the non-thermal physics by means of their synchrotron and $\gamma$-ray
emission. The remnant SN~1006 is an ideal particle acceleration laboratory
because it is observed across all electromagnetic wavelengths from radio to
$\gamma$-rays. We perform three-dimensional (3D) magnetohydrodynamics (MHD)
simulations where we include CR protons and follow the CR electron spectrum. By
matching the observed morphology and non-thermal spectrum of SN~1006 in radio,
X-rays and $\gamma$-rays, we gain new insight into CR electron acceleration and
magnetic field amplification. 1.~We show that a mixed leptonic-hadronic model is
responsible for the $\gamma$-ray radiation: while leptonic inverse-Compton
emission and hadronic pion-decay emission contribute equally at GeV energies
observed by \Fermi, TeV energies observed by imaging air Cherenkov telescopes
are hadronically dominated. 2.\ We show that quasi-parallel acceleration (i.e.,
when the shock propagates at a narrow angle to the upstream magnetic field) is
preferred for CR electrons and that the electron acceleration efficiency of
radio-emitting GeV electrons at quasi-perpendicular shocks is suppressed at
least by a factor ten. This precludes extrapolation of current
one-dimensional plasma particle-in-cell simulations of shock acceleration to
realistic SNR conditions.  3.\ To match the radial emission profiles and the
$\gamma$-ray spectrum, we require a volume-filling, turbulently amplified
magnetic field and that the Bell-amplified magnetic field is damped in the
immediate post-shock region. Our work connects micro-scale plasma physics
simulations to the scale of SNRs.
\end{abstract}




\section{Introduction}
Supernova remnants (SNR) accelerate particles to TeV energies at their shock fronts via diffusive shock acceleration \citep[DSA,][]{Krymskii1977,Axford1977,Blandford1978,Bell1978a, Bell1978b} and are believed to be the source of cosmic rays (CR) in our Galaxy \citep{Reynolds2008}. The remnant of the type Ia supernova SN~1006, also known as the Chinese supernova, is an ideal laboratory to study CR acceleration.
\vspace{2cm}

The shell-type remnant has been observed at various wavebands, e.g. in the radio \citep{Gardner1965, Dyer2009}, infrared \citep{Winkler2013}, optical \citep{Winkler2003}, X-ray \citep{Winkler1997, Bamba2003, Cassam-Chenaie2008, Li2018} and $\gamma$-ray regime \citep{Acero2010, Abdo2010, Condon2017}. It is located approximately \SI{400}{pc} above the Galactic plane within a distance of 1.45 to \SI{2.2}{kpc} \citep{Winkler2003, Katsuda2017}. SN~1006 shows a bilateral symmetry (also called bipolar), i.e. it has radio bright limbs in the northeast (NE) and southwest (SW) which are separated by a dim centre. The location of these spatially coincide with those in non-thermal X- and $\gamma$-rays.

Observations made with the ROSAT and ASCA satellites showed that the X-ray emission at the edges of SN~1006 is due to CR electrons which are accelerated at the shock front and emit synchrotron radiation \citep{Koyama1995, Willingale1996}. The same population of CR electrons is responsible for the radio emission. However, the $\gamma$-ray emission could be a result of CR protons inelastically interacting with the ambient gas (hadronic model) and/or CR electrons scattering off of ambient photons via the inverse Compton (IC) effect (leptonic model). It has been discussed whether the $\gamma$-ray emission of SN~1006 is predominantly of hadronic \citep{Berezhko2012, Miceli2014} or of leptonic origin \citep{Petruk2011, Araya2012, Acero2015, Xing2019}.

The observed morphology in radio and X-rays has been discussed in context of the orientation of the magnetic field and the acceleration mechanism of CR electrons.
In the equatorial-belt model, the magnetic field direction is aligned along the southeast (SE) to northwest (NW) direction and the CR electron acceleration is isotropic or preferentially quasi-perpendicular \citep{Fulbright1990, Reynolds1996, Petruk2009, Schneiter2010}. However, this equatorial-belt model of the magnetic field is in contradiction to the inferred magnetic orientation in radio polarization observations which suggest a magnetic field aligned along the NE-SW direction \citep{Reynoso2013}. This problem is resolved by the polar cap model which relies on a magnetic field oriented along the NE to SW direction and preferentially quasi-parallel acceleration \citep{Voelk2003}. Azimuthal variations of X-ray cutoff frequencies \citep{Rothenflug2004, Katsuda2010} and of the ratio of radii between the forward shock and contact discontinuity \citep{Cassam-Chenaie2008} favour the polar cap model. The polar cap model is further supported by 3D MHD simulations \citep{Bocchino2011, Schneiter2015}.

The observed synchrotron radiation is an indicator of strong magnetic fields. Analysis of the thin X-ray synchrotron rims at SN~1006 suggests post-shock magnetic fields of 70 to \SI{200}{\micro G} \citep{Ressler2014}.
Analysis of the multi-frequency spectrum including the $\gamma$-ray data finds effective (one-zone) magnetic fields of \SI{30}{\micro G} in the case of a leptonic  model and \SI{120}{\micro G} in the case of a hadronic model for the $\gamma$-ray emission \citep{Acero2010}. 

As the remnant SN~1006 evolves in a homogeneous environment high above the Galactic plane, the remnant is surrounded by interstellar magnetic fields of the order of \SI{1}{\micro G}. Therefore, other mechanisms than adiabatic compression of the magnetic fields must be responsible for generating effective magnetic fields with $B \gg \SI{10}{\micro G}$ in the downstream of the shock. First, the non-resonant hybrid instability which is driven by CR protons at the shock amplifies magnetic fields \citep{Bell2004}. Studies of amplified fields at SNRs \citep{Pohl2005} and at relativistic pair plasma shocks \citep{Chang2008, Keshet2009} show that these fields are quickly damped. Secondly, the interaction of the shock with small scale density inhomogeneities of the surrounding interstellar medium can drive a small-scale dynamo which can strongly amplify the magnetic field \citep{Giacalone2007, Ji2016}.

The amplification of magnetic fields is supported by observations of other SNRs, e.g. the variability of X-ray hot spots of the SNR RXJ1713.7-3946 is an indicator of magnetic field amplification up to values larger than \SI{1}{\milli G} \citep{Uchiyama2007}. Another example of highly amplified magnetic fields is the SNR Vela Jr (RXJ0582.0-4622). The analysis of X-ray filaments suggests highly amplified downstream magnetic fields of $B\gtrsim \SI{100}{\micro G}$ which favours a hadronic model for the observed $\gamma$-ray emission \citep{Bamba2005, Berezhko2009a}.
However, a leptonic model with weaker magnetic fields cannot be ruled out \citep{Tanaka2011}
or is favoured if magnetic fields are strongly damped to $\sim\!\SI{10}{\micro G}$ in the downstream of the shock \citep{Sushch2018}.

Here, we study these topics with 3D MHD simulations of the remnant SN~1006 together with magnetic-obliquity dependent acceleration of CR protons and electrons. We follow the spectrum of CR electrons spatially and temporally resolved in order to compare simulations with the observed multi-frequency spectrum and morphology at different wavebands.

Our work has the following structure. We present our simulation setup in Section~\ref{sec:setup}. Then we present our best-fit model and discuss whether the high energy $\gamma$-ray emission is due to leptonic or hadronic processes in Section~\ref{sec:Leptonic_vs_Hadronic}. We continue with the discussion on obliquity dependent acceleration of CR electrons in Section~\ref{sec:Obliquity} and damping of amplified magnetic fields in Section~\ref{sec:damping}. After that, we discuss the influence of various parameters onto the spectrum in Section~\ref{sec:parameters}. We conclude with a discussion of our results in Section~\ref{sec:discussion}. Throughout this work, we denote photon energies by $E$, electron energies and normalised (dimensionless) momenta by $E_\rmn{e}$ and $p_\rmn{e} = P_\mathrm{e} / (m_\mathrm{e} c)$, and proton energies and normalised momenta by $E_\rmn{p}$ and $p_\rmn{p} = P_\mathrm{p}/(m_\mathrm{p} c)$. Here, $P_\mathrm{e}$ and $P_\mathrm{p}$ are the physical electron and proton momenta in units of \si{\gram\centi\meter\per\second}, respectively, $m_\mathrm{e}$ and $m_\mathrm{p}$ are the electron and proton masses, respectively, and $c$ is the speed of light.

\section{Simulation Setup}
\label{sec:setup}
\begin{figure}
	\includegraphics[width=\columnwidth]{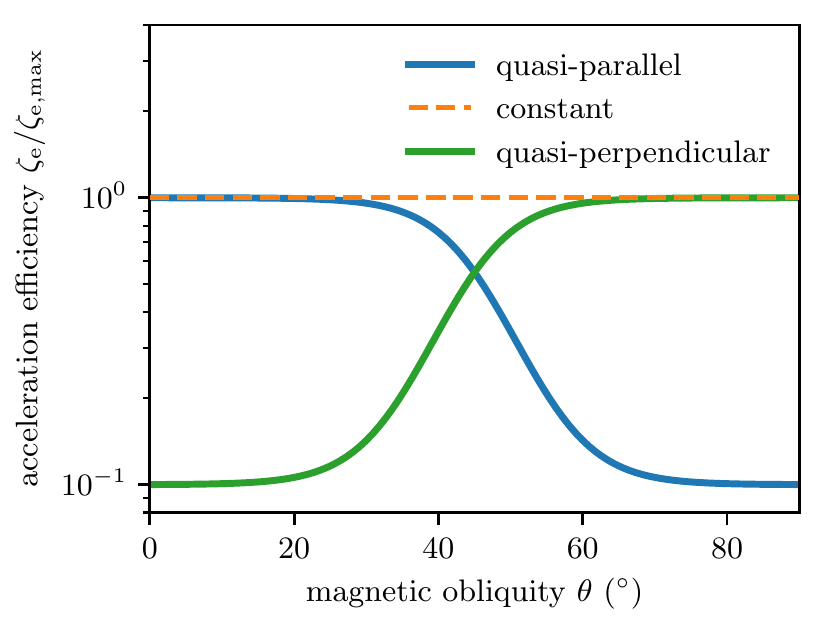}
	\caption{Obliquity dependent electron acceleration efficiency for three base acceleration models. Best fits are produced with the preferred quasi-parallel acceleration efficiency (blue).}
	\label{fig:Obliquity_Efficiency}
\end{figure}
\subsection{Simulation codes}

We perform 3D MHD simulations with the second-order
accurate, adaptive moving-mesh code \textsc{arepo} \citep{Springel2010,
Pakmor2016} which employs an unstructured mesh that is defined as the Voronoi
tessellation of a set of mesh-generating points. We account for the transport of
CR protons which are treated as a relativistic fluid with an effective adiabatic
index of 4/3 \citep{Pfrommer2017a}. We employ the shock
finder \citep{Schaal2015} which localises and characterises shocks according to
the Rankine--Hugoniot jump conditions and inject CR energy into the Voronoi
cells of the shock and immediate post-shock regime \citep{Pfrommer2017a}.  We
account for the dominant advective transport of CR protons and neglect CR
streaming and diffusion. While a combination of adiabatic gains due to the
converging flow at the shock and spatial diffusion (close to the Bohm limit)
gives rise to diffusive shock acceleration \citep{Blandford1987}, we do not
resolve the growth of non-resonant \citet{Bell2004} modes of the hybrid
instability in our simulations. In consequence, we describe diffusive shock
acceleration as well as Bell amplification in form of subgrid models detailed
below in Sections~\ref{sec:acc} and \ref{sec:ICs}.

In addition, we follow the evolution of the CR electron spectrum spatially and
temporally resolved in post-processing with the \textsc{crest}
code \citep{Winner2019}. CR electrons are evolved according to the
Fokker--Planck equation on Lagrangian tracer particles while taking adiabatic
changes and cooling via Coulomb losses, synchrotron emission, and IC processes
into account. If tracer particles encounter the shock front, we model Fermi I
acceleration via injecting a fraction of the dissipated thermal energy into
a momentum power-law spectrum with logarithmic slope
$\alpha_\mathrm{e}$. As a result, we obtain the CR proton energy density and
the CR electron momentum spectrum at every time and at every point in our 3D
simulation domain.

\subsection{Magnetic obliquity-dependent acceleration}
\label{sec:acc}
\begin{figure*}
	\includegraphics[width=\textwidth]{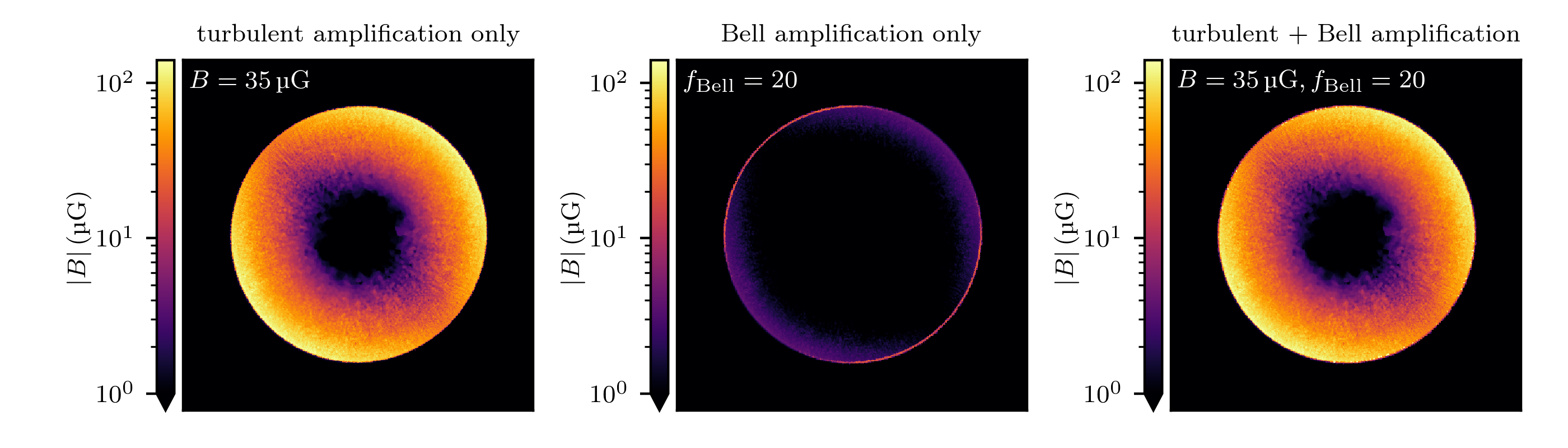}
	\caption{Magnetic field morphology in a slice through the centre of our simulated remnant. The left-hand panel shows the magnetic field strength for turbulent amplification only, the central panel shows the effect of Bell amplification only, and the right-hand panel shows the magnetic field strength as result of both, turbulent and Bell amplification. The maps have a side length of \SI{21}{pc} or \ang{;42.5;} at a distance of \SI{1660}{pc}.
	}
	\label{fig:Morphology_MagneticField}
\end{figure*}
The efficiency of proton acceleration, for quasi-parallel shocks (with magnetic
field aligned close to the shock normal) can be up to 10 to 20 per cent of the initial
shock kinetic energy, and the efficiency drops to zero for quasi-perpendicular
shocks \citep{Caprioli2014}, which is a consequence of the effective excitation
of the non-resonant hybrid instability at quasi-parallel shocks that enables
efficient CR proton acceleration \citep{Caprioli2014a,Caprioli2015}. We thus
adopt an acceleration efficiency for CR protons that depends on the upstream
magnetic obliquity $\theta$, the angle between the shock normal and the upstream
magnetic field according to \citep{Pais2018}
\begin{align}
\zeta_\rmn{p} (\theta) = \frac{\zeta_\mathrm{p, max}}{2}  \left[\tanh(\frac{\theta_\mathrm{crit} - \theta}{\delta}) + 1 \right]
\end{align}
where $\zeta_\mathrm{p, max}=0.15$ is the efficiency in quasi-parallel
configurations, $\theta_\mathrm{crit} = \pi/4$ the critical obliquity, and
$\delta = \pi/18$ the shape parameter. We define the acceleration efficiency
as the ratio of accelerated CR energy density to the total shock-dissipated
energy density, $\zeta_\rmn{p,e}=\varepsilon_{\rmn{CRp,e}}
/ \varepsilon_{\rmn{diss}}$. 

Following the algorithm described in \citet{Winner2019}, we inject a CR electron
spectrum once a Lagrangian tracer particle crosses the shock and use the
parametrized form of the 1D CR electron momentum spectrum (in units of particles
per unit volume)
\begin{align}
f_\mathrm{e}(p_\rmn{e},\theta) &= C_\mathrm{e}(\theta) p_\rmn{e}^{-\alpha_\mathrm{e}}
\Theta(p_\rmn{e} - p_\mathrm{inj})\,\nonumber\\
 &\times\left[1 + a \left(\frac{p_\rmn{e}}{p_\mathrm{cut}}\right)^b \right]^c \exp[
 - \left(\frac{p_\rmn{e}}{p_\mathrm{cut}}\right)^2],
\end{align}
where we adopt the parameters $a=0.66$, $b=2.5$, and $c=1.8$ and treat the
electron spectral index $\alpha_\mathrm{e}$ and the (normalised) cutoff momentum
$p_\mathrm{cut}$ as free parameters that we vary in this work. 

The normalisation $C_\mathrm{e}$ and injection momentum $p_\mathrm{inj}$
are calculated for every Lagrangian particle by attaching the non-thermal power-law spectrum to a thermal
Maxwellian. We require that the energy moment of the distribution function equals the CR electron energy
density,
\begin{align}
\varepsilon_\mathrm{CRe}= m_\mathrm{e} c^2 \int_0^\infty f_\rmn{e}(p_{\rmn{e}})
\left[\sqrt{1 + p_{\mathrm{e}}^2} - 1\right]\,  \mathrm{d}p_{\rmn{e}},
\end{align}
which we compare to the dissipated energy density at the shock according to our
specific model of obliquity dependent shock acceleration that we describe now.

The acceleration efficiency of CR electrons $\zeta_\mathrm{e}$ depends on the magnetic obliquity angle $\theta$
\begin{align}
\zeta_\rmn{e} (\theta) = \frac{\zeta_{\rmn{e}, \parallel} - \zeta_{\rmn{e}, \perp}}{2}  \left[\tanh(\frac{\theta_\rmn{crit} - \theta}{\delta}) + 1 \right] + \zeta_{\rmn{e}, \perp} \label{eq:obliquity}
\end{align}
where $\zeta_{\rmn{e}, \parallel}$ is the quasi-parallel acceleration efficiency for $\theta=0$ and $\zeta_{\rmn{e}, \perp}$ is the quasi-perpendicular efficiency for $\theta=\pi/2$ (i.e., for $90^\circ$). Ab initio, the functional form of equation~\eqref{eq:obliquity} is not known. Thus we explore three different models that are motivated by different lines of physics arguments and confront them to observational data.

One-dimensional (1D) particle-in-cell simulations of non-relativistic, high
Mach number, {\em quasi-parallel} shocks \citep{Park2015} find the onset of
acceleration of non-thermal electrons and protons, in agreement with the
predictions of the theory of diffusive shock acceleration. On the other hand,
full particle-in-cell simulations show indications that electrons may be
possibly even more efficiently accelerated at {\em quasi-perpendicular},
high-Mach number shocks \citep{Riquelme2011,Bohdan2017,Xu2020}. The electron
acceleration efficiency is 0.1 by energy relative to the downstream thermal
electrons \citep{Xu2020}, which have a fraction of 0.1 of the energy of the
downstream thermal protons (Spitkovsky, private comm.). 
Combining this, we
obtain an overall acceleration efficiency relative to the dissipated energy of
about $\varepsilon_\rmn{CRe} / \varepsilon_\rmn{diss} \approx 10^{-2}$ for {\em
quasi-perpendicular} strong shocks. By contrast, the electron acceleration
efficiency of {\em quasi-parallel} strong shocks is $\varepsilon_\rmn{CRe}
/ \varepsilon_\rmn{diss} \lesssim 10^{-3}$ \citep{Park2015, Xu2020}.

This motivates our {\em quasi-perpendicular acceleration model}, for which
we assume $\zeta_{\mathrm{e}, \perp}=10\zeta_{\mathrm{e}, \parallel}$. This would
be the correct model provided we can extrapolate the short simulation time of
physical seconds to the SNR live
time of more than 1000 years and provided there are no multi-dimensional
effects that interfere with the extrapolations of these 1D particle-in-cell
simulations. We contrast this model with two alternative models: in our {\em
quasi-parallel acceleration model}, we assume $\zeta_{\mathrm{e}, \parallel} =
10\zeta_{\mathrm{e}, \perp}$ and in a third model we adopt a {\em constant
acceleration efficiency}, $\zeta_e(\theta) = \zeta_\mathrm{e, max}$.

The maximum acceleration efficiency of CR electrons
$\zeta_\mathrm{e,max}=\rmn{max}(\zeta_{\mathrm{e}, \parallel}, \zeta_{\mathrm{e}, \perp})$
is a free parameter which is set such that a spectral fit to radio data is
obtained. We obtain values of $\zeta_\mathrm{e, max} < 10^{-3}$ which reflect
that the ratio of electron-to-proton acceleration efficiency is
$\zeta_\mathrm{e, max} / \zeta_\mathrm{p, max}<10^{-2}$ \citep{Schlickeiser2002, Zweibel2013}. 
The obliquity dependency of quasi-parallel, constant, and
quasi-perpendicular acceleration models are shown in
Figure~\ref{fig:Obliquity_Efficiency}.

\subsection{Initial conditions}
\label{sec:ICs}
\begin{table}
	\caption{Resolution per box length (\SI{22}{pc}) in shells from the centre. The low resolution contains $10^6$ cells and the high resolution $5\times 10^6$ cells.}
	\begin{tabular}{rrr}
		\hline
		radius $r$ (pc) & low resolution & high resolution \\
		\hline\noalign{\smallskip}
		\num{3.1} & 300 & 300 \\
		\num{6.2} & 200 & 250 \\
		\num{9.3} & 100 & 200 \\
		\num{12.4}& 75 & 150 \\
		\num{15.6} & 50 & 100 \\		
		\hline
	\end{tabular}
	\label{tab:Resolution}
\end{table}
In order to  model the remnant of the Type Ia SN~1006, we inject \SI{e51}{erg} of thermal energy into the central cell of a periodic 3D box with \SI{22}{pc} length. We use two setups, the first with a resolution of $10^6$ cells for parameter space studies and the second with a high resolution of $5\times 10^6$ cells for morphological studies. The cells are distributed in five shells around the centre and the average cell density per box length decreases from the first to the last shell as shown in Table~\ref{tab:Resolution}. The centres of the cells are then perturbed by 10 per cent of the local average cell length before we relax the mesh via Lloyd's algorithm \citep{Lloyd1982}  in order to obtain glass-like configurations. We chose a higher cell density in the centre of our simulation box because of the fast initial adiabatic expansion of this central region. Tracer particles are initially sampled on positions of the cell centres except for a small exclusion region within a radius \SI{0.55}{pc} around the centre due to high numerical noise before the shock has developed numerically over a few cells.

As initial conditions, we adopt a gas number density of ${n=\SI{0.12}{cm^{-3}}}$, a mean molecular weight of $\mu=1.25$, and temperature of $T=\SI{5.1e3}{K}$. The initial magnetic field is oriented along the diagonal of the plane of the sky and has an absolute value of $B=\SI{1}{\micro G}$.
This setup leads to an energy driven, spherical shock wave driving into a homogeneous medium. We ignore the free expansion phase of the remnant as it's influence onto the final radius is smaller than 10 per cent \citep{Pais2020}.

\subsection{Magnetic modelling}
The magnetic field in the simulation is affected by three physical processes. First, the adiabatic compression at the shock enhances the magnetic field. However, only the component perpendicular to the shock normal is amplified by a factor $n_\mathrm{post}/n_\mathrm{pre}$ where $n_\mathrm{pre}$ and $n_\mathrm{post}$ are the pre- and post-shock gas densities, respectively.

Secondly, we model the effect of a turbulent dynamo that is generated as a result of the interaction between pre-shock turbulence, clumping and the shock \citep{Ji2016} which leads to high post-shock fields. Throughout our work, we multiply the magnetic field of our MHD solution inside the SNR and behind the shock front by an amplification factor and refer to an equivalent magnetic field strength instead of the amplified field (which differs for quasi-parallel and -perpendicular shock morphologies). Hence, a field of $B=\SI{35}{\micro G}$ is equivalent to a turbulent amplification of the post-shock fields by a factor of 35 for the parallel shock configuration and reaches a field strength of $\SI{140}{\micro G}$ for the perpendicular shock configuration in the equatorial region.

Thirdly, we employ the amplification of magnetic fields via the non-resonant hybrid instability which is driven by the CR proton current in the pre-shock region \citep{Bell2004}. This so called Bell amplification drives strong perpendicular magnetic fields that are responsible for the efficient acceleration of CR protons (and possibly also CR electrons) in the quasi-parallel regime. We compute a cell-averaged value of the amplified field with
\begin{align}
\abs{B_\mathrm{amp}} = \abs{B} \left( \frac{f_\mathrm{Bell} - 1}{2}  \left[\tanh(\frac{\theta_\mathrm{crit} - \theta}{\delta}) + 1 \right] + 1\right)
\end{align}
which follows the obliquity dependency of CR proton acceleration. We parametrize the Bell amplification by an amplification factor $f_\mathrm{Bell}$ which can reach values of about 30 \citep{Bell2004}.

We note that both amplification processes hardly overlap as the Bell amplified fields are quickly damped in the post-shock region whereas the turbulently amplified magnetic field starts to build up in the post-shock region as the small-scale dynamo emerges but saturates only after a finite time and distance from the shock.

Figure~\ref{fig:Morphology_MagneticField} shows the resulting magnetic morphology in a slice through the centre of our simulated remnant. In the left-hand panel, the magnetic field is only amplified by the turbulent dynamo (and by adiabatic compression) to values of $B= 35$ to $\SI{140}{\micro G}$ for quasi-parallel and -perpendicular geometries, respectively. The shock front encounters small-scale density inhomogeneities which inject vorticity according to \citeauthor{Crocco1937}'s theorem (\citeyear{Crocco1937}) that leads to a turbulent cascade and a small-scale dynamo, which amplifies the post-shock magnetic field. This process saturates if the magnetic energy density reaches about 10 per cent of the kinetic energy density in the post shock medium \citep{Schekochihin2004, Cho2009, Kim2015, Federrath2016}. In the post-shock rest frame, the shock velocity is $\varv = 3/4 \varv_\mathrm{s}$, where $\varv_\mathrm{s}$ is the lab-frame shock velocity. Hence the maximum possible value of the turbulently amplified magnetic field (neglecting adiabatic cooling) is given by
\begin{align}
B_\mathrm{max, turb} \approx \sqrt{0.1 \times 8 \pi \frac{4 \mu m_\mathrm{p} n }{2} \left(\frac{3}{4}\varv_\mathrm{s} \right)^2 } \approx \SI{250}{\micro G}, \label{eq:Bell_Amplification}
\end{align}
where $n=\SI{0.12}{\cm}^{-3}$ is the pre-shock density and $\varv_\mathrm{s} \approx \SI{3000}{km \per \second}$ is the shock velocity 1000 years after explosion. This is larger than the \SI{140}{\micro G} field that our best-fit model requires in the quasi-perpendicular regions. In the quasi-parallel regions, a lower magnetic field of $\SI{35}{\micro G}$ is realised as the turbulent dynamo acts on magnetic seed values that are four times smaller due to the absence of adiabatic compression for the parallel field geometry, but saturates at the same time as for the quasi-perpendicular regions.

In the central panel of Figure~\ref{fig:Morphology_MagneticField}, we show the magnetic morphology for Bell amplification only. Bell amplification scales with the magnetic obliquity due to obliquity dependent acceleration of CR protons as shown in Equation~\eqref{eq:Bell_Amplification} such that it steeply declines towards quasi-perpendicular regions. Due to its small spatial extend at the shock front, the Bell amplified magnetic field barely influences the cooling of the CR electron spectrum. Note that the field strengths of approximately \SI{4}{\micro G} in the quasi-perpendicular regions are solely due to adiabatic compression.

The right-hand panel of Figure~\ref{fig:Morphology_MagneticField} shows the magnetic field as a result of both amplification processes. As explained before, the magnetic field is dominated by the turbulent amplification and the Bell-amplified field plays a minor role for the overall magnetic morphology.

\subsection{Non-thermal radiative transfer}
\begin{figure*}
	\includegraphics[width=\textwidth]{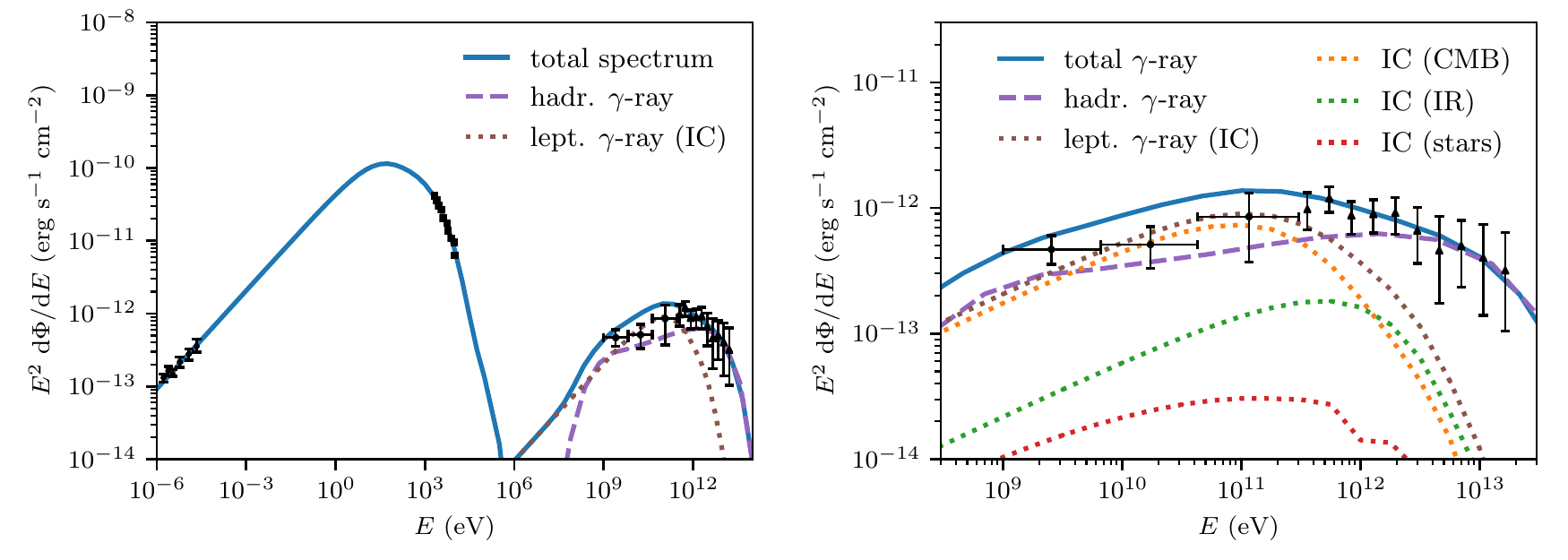}
	\caption{Best fit multi-frequency spectrum of SN~1006. The CR spectra have spectral indices of $\alpha_\mathrm{e}=2.1$ for electrons and $\alpha_\mathrm{p}=1.9$ for protons. Important parameters are gas density of $n=\SI{0.12}{\per\cubic \centi\meter}$, distance of $D=\SI{1660}{pc}$, equivalent magnetic field of $B=\SI{35}{\micro G}$ (as a result of a turbulent dynamo), and Bell amplification by a factor of 20. The simulated spectrum is compared with observational data in radio \citep{Reynolds1996}, X-rays \citep{Bamba2008}, and in $\gamma$-rays from \Fermi \citep{Abdo2010} and HESS \citep{Acero2010} (sum of two regions).}
	\label{fig:BestFit_Spectrum}
\end{figure*}
Non-thermal synchrotron and IC emission is calculated from the simulated CR electron spectra. 
We assume an isotropic distribution of pitch angles for synchrotron emission
and follow the analytic approximation by \citet{Aharonian2010}. For the IC emission, we include the Klein-Nishina cross section \citep{Blumenthal1970}.
In contrast to CR electrons, the simulations evolve only the energy density of CR protons $\varepsilon_\mathrm{CRp}$. In order to calculate hadronic $\gamma$-ray emission, we calculate a 1D CR proton spectrum of the form
\begin{align}
f_\mathrm{p}(p_{\rmn{p}}) = C p_{\rmn{p}}^{-\alpha_\mathrm{p}} \Theta(p_{\rmn{p}} - q) \exp\left[-\left(\frac{p_{\rmn{p}}}{p_\mathrm{max}}\right)^2\right], \label{eq:ProtonSpectrum}
\end{align}
where $\alpha_\mathrm{p}$ is the logarithmic momentum slope,  $q=0.5$ is the minimum momentum, and $p_\mathrm{max} = \num{2.1e5}$ is the (normalised) maximum momentum. The normalisation $C$ is calculated for every cell such that the energy moment of the distribution function equals the proton energy density, $\varepsilon_\mathrm{CRp}= m_\mathrm{p} c^2 \int f_\mathrm{p}(p_{\rmn{p}}) [(1 + p_{\rmn{p}}^2)^{1/2} - 1]\, \mathrm{d}p_{\rmn{p}}$. Hadronic gamma ray emission is then obtained with parametrizations of the cross-section of neutral pion production at low ($E_\mathrm{p, kin} < \SI{10}{GeV}$) and high proton energies (\citealt{YangRz2018, Kafexhiu2014}, Werhahn et al.\ in prep.).

The synthetic noise map is based on the noise power spectrum of the excess map of SN~1006. To detect the noise, we exclude the emission from the NE and SW lobes masking the original excess map from \citet{Acero2010} with a sharp cutoff calculated taking the absolute value of the minimum of the excess counts. The power spectrum of SN~1006 is obtained via a 2D Fourier transform of the masked data set. We fit the power spectrum with the following function in k-space
\begin{align}
P(k) = A \exp(- \frac{k^2}{2 \sigma^2}) + B k^{-2}
\end{align}
where $\sigma_k$ is the standard deviation in $k$-space and the variables $A$ and $B$ determine the relative strength of the Gaussian and the power-law tail. The fitted power spectrum is converted into a real noise map via 2D inverse Fourier transform and added in post processing to the previously PSF-convolved simulation map.

\section{Leptonic versus hadronic model}
\label{sec:Leptonic_vs_Hadronic}

\begin{figure*}
	\includegraphics[width=\textwidth]{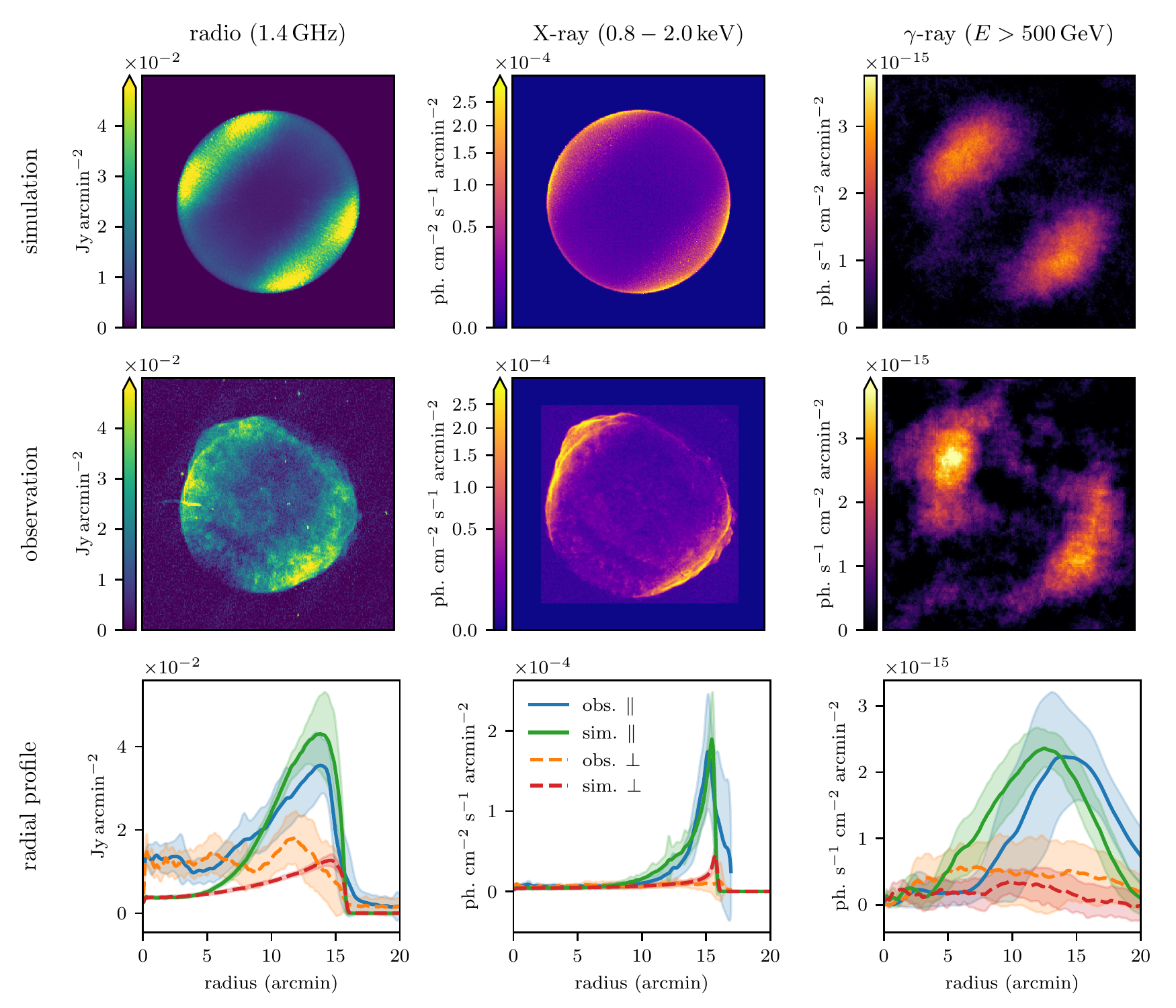}
	\caption{Morphological comparison of our best-fit simulation with preferred {\em quasi-parallel acceleration} to \SI{1.4}{GHz} radio data \citep{Dyer2009} (left column), to \SIrange{2}{4.5}{keV} X-ray data \citep{Cassam-Chenaie2008} (middle column), and to HESS $\gamma$-ray data above \SI{500}{GeV} \citep{Acero2010} (right column). 
	The top row shows our simulated maps and in the middle row observations are presented. In the bottom row, we show radial profiles for data in sectors with opening angle of $\pi/3$ aligned parallel and perpendicular to the magnetic field. Our simulated $\gamma$-ray map is convolved with a 2D Gaussian profile with $\sigma = 0.042^\circ$ similar to the HESS PSF and contains Gaussian noise with the observed amplitude and correlation structure (as quantified through the power spectrum where we cut the signal regions). The maps have a side length of \SI{21}{pc} or \ang{;42.5;} at a distance of \SI{1660}{pc}.
	}
	\label{fig:Morphology_Parallel}
\end{figure*}

\begin{figure}
	\includegraphics[width=\columnwidth]{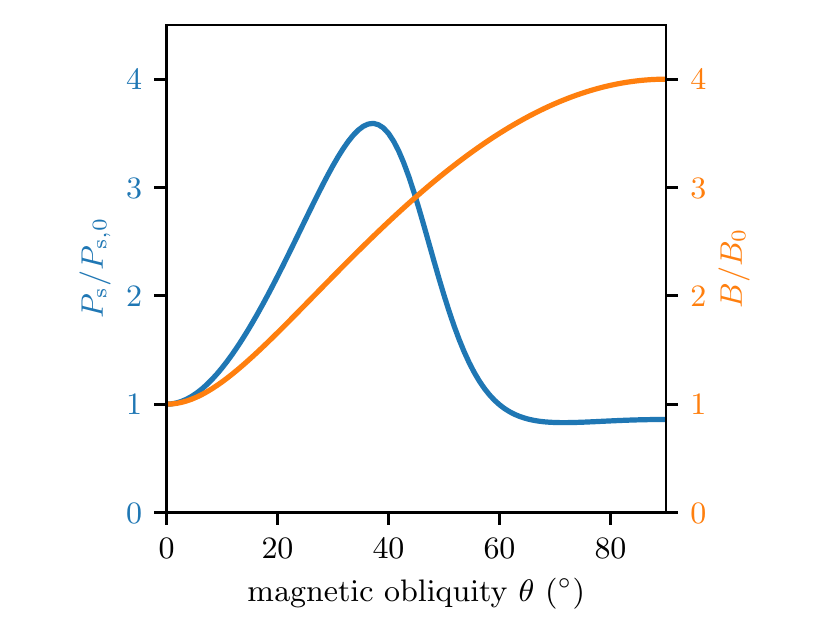}
	\caption{Angular dependency of synchrotron emission (blue) and adiabatically compressed magnetic field (orange) relative to their values at $0^\circ$ for our best-fit quasi-parallel acceleration model.}
	\label{fig:Obliquity_Magnetic_Field}
\end{figure}

\begin{figure*}
	\includegraphics[width=\textwidth]{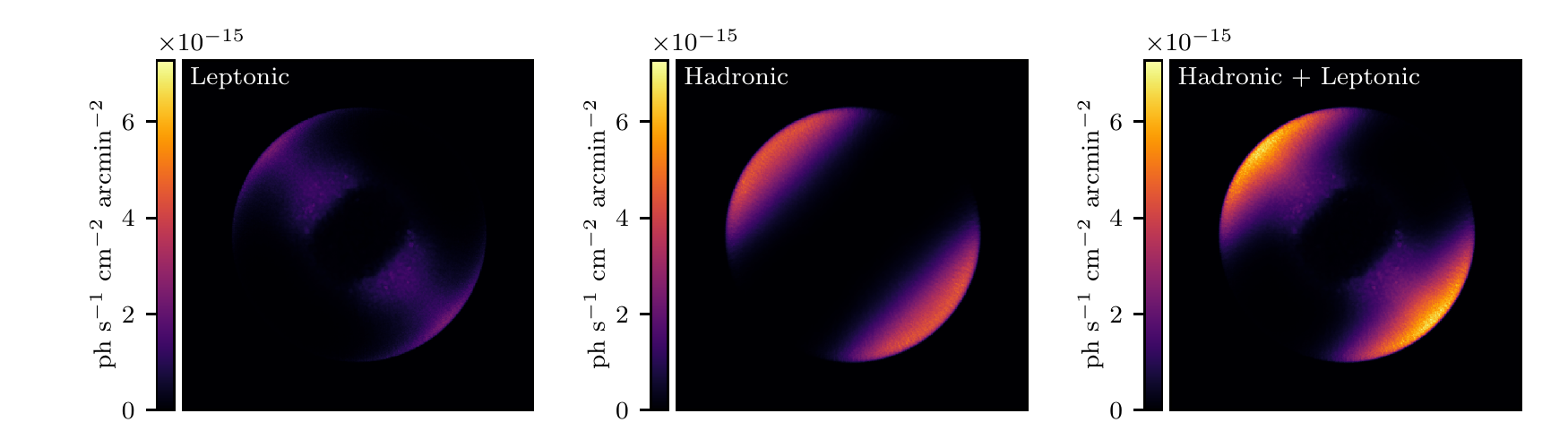}
	\caption{$\gamma$-ray maps ($E > \SI{500}{GeV}$) of our best-fit quasi-parallel acceleration model. We compare the leptonic IC emission (left), the hadronic  pion-decay emission (middle) and the total $\gamma$-ray emission (right), which is dominated by the hadronic component. The maps have a side length of \SI{21}{pc} or \ang{;42.5;} at a distance of \SI{1660}{pc}.}
	\label{fig:Morphology_Gamma_Ray}
\end{figure*} 

In this section, we present our best-fit simulation together with its multi-frequency spectrum and maps of radio, X-ray, and $\gamma$-ray surface brightness. We compare these to observations and discuss whether leptonic or hadronic emission is dominating in the high energy $\gamma$-ray regime. 

\subsection{Multi-frequency spectrum}

In Figure~\ref{fig:BestFit_Spectrum}, we present the multi-frequency spectrum of our best-fit simulation. The simulation uses a homogeneous gas density of $n=\SI{0.12}{cm^{-3}}$, an equivalent magnetic field of $B=\SI{35}{\micro G}$, a Bell amplification by the factor 20 at the shock front, a CR electron spectral index of $\alpha_\mathrm{e}=2.1$, and a maximum CR electron acceleration momentum of $p_\mathrm{cut} = \num{3.5e7}$. In our best-fit model, CR electron acceleration is most efficient in quasi-parallel configurations. We discuss how variations of these parameters or prescriptions impact the spectrum or the emission morphologies in Sections~\ref{sec:Obliquity} to \ref{sec:parameters}.

The radius of our simulated remnant and the observed angular size of \ang{;0.5;} yields a distance to the remnant of $D=\SI{1660}{pc}$. We leave the electron acceleration efficiency $\zeta_\mathrm{e,max}$ as a free parameter in order to fit the observed radio data. The spectrum fits the data very well with an acceleration efficiency of $\zeta_\mathrm{e,max} = \num{5e-4}$. The synchrotron spectrum has a spectral index of $\alpha_\mathrm{s}= (\alpha_\mathrm{e} -1)/2 = 0.55$ up to photon energies of $E \approx \SI{10}{eV}$. At larger photon energies the synchrotron spectrum is sensible to the cooling of the underlying CR electron spectrum and its cutoff. The dominant electron momentum\footnote{We obtain this formula by replacing the kernel $F(E/E_c)$ in the synchrotron emissivity by Dirac's $\delta$ distribution at its expectation value $E / E_c = 4 \pi m_\mathrm{e} c \nu / (3 e B ) \approx 2.13$, e.g., see equation~(D1) in \citet{Aharonian2010}.} for emission at synchrotron frequency $\nu_\mathrm{s}$ is
\begin{align}
p_\rmn{e}\approx 5\times10^3\,\left(\frac{\nu_\mathrm{s}}{\SI{1}{GHz}}\right)^{1/2} \left(\frac{B}{\SI{5}{\micro G}}\right)^{-1/2}. \label{eq:Synchrotron_Dominant_Momentum}
\end{align}
 Hence, the dominant momentum for \SI{1}{keV} X-rays at \SI{35}{\micro G} is $p_\rmn{e}\approx \num{3e7}$ which is close to the maximal electron acceleration momentum. This explains the synchrotron cutoff at X-ray energies.

At even larger photon energies, in the GeV to TeV $\gamma$-ray range, the photon spectrum is a combination of leptonic emission from IC and hadronic $\gamma$-ray emission from CR protons interacting with the ambient gas. We assume that the IC emission results from CRe interactions with three black-body photon fields: the cosmic microwave background (CMB), an infrared field with $T_\mathrm{IR} = \SI{30}{K}$, and a star light photon field $T_\mathrm{star} = \SI{4100}{K}$ (see Section~\ref{sec:parameters} for details of the adopted radiation fields). Leptonic emission is dominating over hadronic emission at \Fermi $\gamma$-ray energies from 1 to \SI{100}{GeV}. For photon energies larger than \SI{100}{GeV}, the IC spectrum falls off as it is influenced by the maximal momentum of the underlying CR electrons. Hadronic $\gamma$-ray emission is therefore dominating at very-high $\gamma$-ray energies observed by the High Energy Stereoscopic System (HESS).

\subsection{Non-thermal emission morphologies}

In Figure~\ref{fig:Morphology_Parallel}, we compare simulated and observed morphology of SN~1006.  We present three simulated surface brightness maps of radio, X-ray, and $\gamma$-ray emission (top row) together with the corresponding images from observations (middle row). Observational images are rescaled such that the integrated surface brightness corresponds to the spectral data. In addition, we show radial profiles for regions quasi-parallel and quasi-perpendicular to the magnetic field (bottom row). Radial profiles are created by selecting sectors of size $\pi/3$ around the magnetic field vector (NE to SW direction) and around the perpendicular vector to the magnetic field in the plane of the sky (SE to NW direction).

Our simulated radio map (top left panel of Figure~\ref{fig:Morphology_Parallel}) matches well the observed map of SN~1006 (mid left panel). It is a combination of single dish observations with the Green Bank Telescope and interferometric observations with the Very Large Array at \SI{1.4}{GHz} \citep{Dyer2009}. The map shows bright polar caps in the NE and SW direction and regions of low surface brightness in the centre, the NW and SE direction.\footnote{The bright elongated source in the eastern rim is a background radio galaxy \citep{Cassam-Chenaie2008}.} The polar caps are bright due the efficient acceleration of CR electrons in these regions with quasi-parallel shock acceleration. Regions with low surface brightness are characterised by an acceleration efficiency of CR electrons that is smaller by a factor of 10 (see Figure~\ref{fig:Obliquity_Efficiency}) due to the quasi-perpendicular shock morphology.

This comparison is quantified through the radial profiles for the quasi-parallel and quasi-perpendicular regions (bottom left panel), demonstrating a very good agreement except for the central regions which show a slightly elevated emission in the observations. After acceleration at the shock, the CR electrons are advected downstream, cool adiabatically and suffer radiation losses so that the central region of the remnant experiences low radio synchrotron surface brightness. Our simulations do not explicitly account for a turbulent dynamo and may thus underestimate the level of magnetic fluctuations inside the SNR. There may even be reacceleration of CR electrons at magnetic reconnection sites or by interacting with the MHD turbulence that counteracts some of the CR electron cooling processes.

Each polar cap shows two bright spots at an angle of $\theta \approx 35^\circ$ that exceed the emission at the parallel orientation of $\theta \approx 0^\circ$. The reason for this is a competition of two effects that have different azimuthal dependencies. At quasi-perpendicular shock morphologies, the ambient magnetic field is adiabatically compressed by a factor of four at the shock front and remains unaltered at quasi-parallel shock morphologies. Our quasi-parallel acceleration efficiency (see equation~\ref{eq:obliquity}) shows the opposite behaviour and peaks at quasi-parallel morphologies. It turns out that the adiabatic magnetic field amplification increases faster with the increasing obliquity angle than the acceleration efficiency decreases, which results in the particular azimuthal behaviour of the radio surface brightness that is shown in Figure~\ref{fig:Obliquity_Magnetic_Field}.

We draw similar conclusions from the comparison between observation and simulation of the X-ray surface brightness map (central column of Figure~\ref{fig:Morphology_Parallel}) for \SIrange{0.8}{2}{keV} photons. The simulated X-ray map (top central panel) has a similar morphology in comparison to the simulated radio map. Polar caps are visible which are a consequence of the efficient acceleration of CR electrons in quasi-parallel regions where the magnetic field is parallel to the shock normal. The emission in the polar caps also peaks at around an angle of $\theta \approx 35^\circ$ away from the magnetic field axis. Regions where the magnetic field is perpendicular to the shock normal have a lower surface brightness.

The simulated X-ray map shows rims contrary to the simulated radio map where the emitting regions shows a larger extend towards the centre. This is because the CR electron momentum that emits X-ray synchrotron emission (see equation~\ref{eq:Synchrotron_Dominant_Momentum}) is close to the maximum acceleration momentum $p_\mathrm{cut} = \num{3.5e7}$. These CR electrons cool fast by means of synchrotron emission in strong magnetic fields and by adiabatic expansion as the SNR expands. Therefore, the spectrum at electron momenta relevant for X-ray emission plummets towards the centre. Strong non-thermal X-ray emission is therefore only present at the shock front where CR electrons are freshly accelerated to the X-ray synchrotron emitting momentum. The simulated X-ray map matches the observed X-ray map (middle centre panel of Figure~\ref{fig:Morphology_Parallel}) which is processed from \textit{Chandra} observations \citep{Cassam-Chenaie2008}. The radial profiles (bottom centre panel) show again excellent agreement between simulation and observation in quasi-parallel and quasi-perpendicular regions.

In the right column, we compare simulation and observation in the $\gamma$-ray band above $E_\mathrm{\gamma} > \SI{500}{GeV}$.
The simulated $\gamma$-ray map (top right panel of Figure~\ref{fig:Morphology_Parallel}) is a sum of leptonic and hadronic $\gamma$-ray emission. The map is convolved with a 2D Gaussian profile with $\sigma=0.042^\circ$ similar to the HESS point spread function (PSF).\footnote{The HESS PSF has a 68 per cent containment radius of $R_\mathrm{68}~=~0.064^\circ$ \citep{Acero2010}. This corresponds to $\sigma \approx R_{68} / 1.515$ for a 2D Gaussian profile \citep{Stycz2016}.} In addition, we add Gaussian noise with the observed amplitude and correlation structure (as quantified through the power spectrum where we cut the signal regions). The map shows two bright, elongated emission regions tracing out a quasi-parallel shock morphology. These regions spatially coincide with those in the radio and X-ray. However, no emission is visible in the centre and in the quasi-perpendicular regions in contrast to the radio and X-ray maps. 
The morphology of the simulated $\gamma$-ray map matches that of the observed map (middle right panel of Figure~\ref{fig:Morphology_Parallel}). Observations were made with HESS and analysed by \cite{Acero2010}. The radial profiles (bottom right panel) show very good agreement between our simulation and the observation.

In Figure~\ref{fig:Morphology_Gamma_Ray}, we show the $\gamma$-ray maps of leptonic and hadronic emission as well as the sum of both processes. Leptonic emission (left panel) results from IC interactions with three photon fields, of which the IC emission from CMB photons is dominant. IC emission produces thin rims in the quasi-parallel regions where the CR electron efficiency is at its maximum. There are tails of IC emission parallel to the magnetic field towards the centre because the CR electron spectrum is less affected by synchrotron cooling in comparison to the regions of larger obliquity angles where the magnetic field is compressed adiabatically. In our model, quasi-perpendicular regions do not shine via IC emission as the CR electron acceleration efficiency there is lower by a factor of 0.1 in comparison to the quasi-parallel region.

Hadronic emission (centre panel of Figure~\ref{fig:Morphology_Gamma_Ray}) is calculated from the decay of neutral pions resulting from the interaction of CR protons with the protons of the gas. CR protons are accelerated efficiently in quasi-parallel regions whereas the efficiency drops to zero for quasi-perpendicular regions. Therefore, hadronic processes produce extended, bright polar caps in $\gamma$-rays. There is no hadronic $\gamma$-ray emission towards the centre as CR protons cool adiabatically and the target gas density is decreasing as a power law in radius.

\section{Obliquity dependent acceleration}
\label{sec:Obliquity}
\begin{figure*}
	\includegraphics[width=\textwidth]{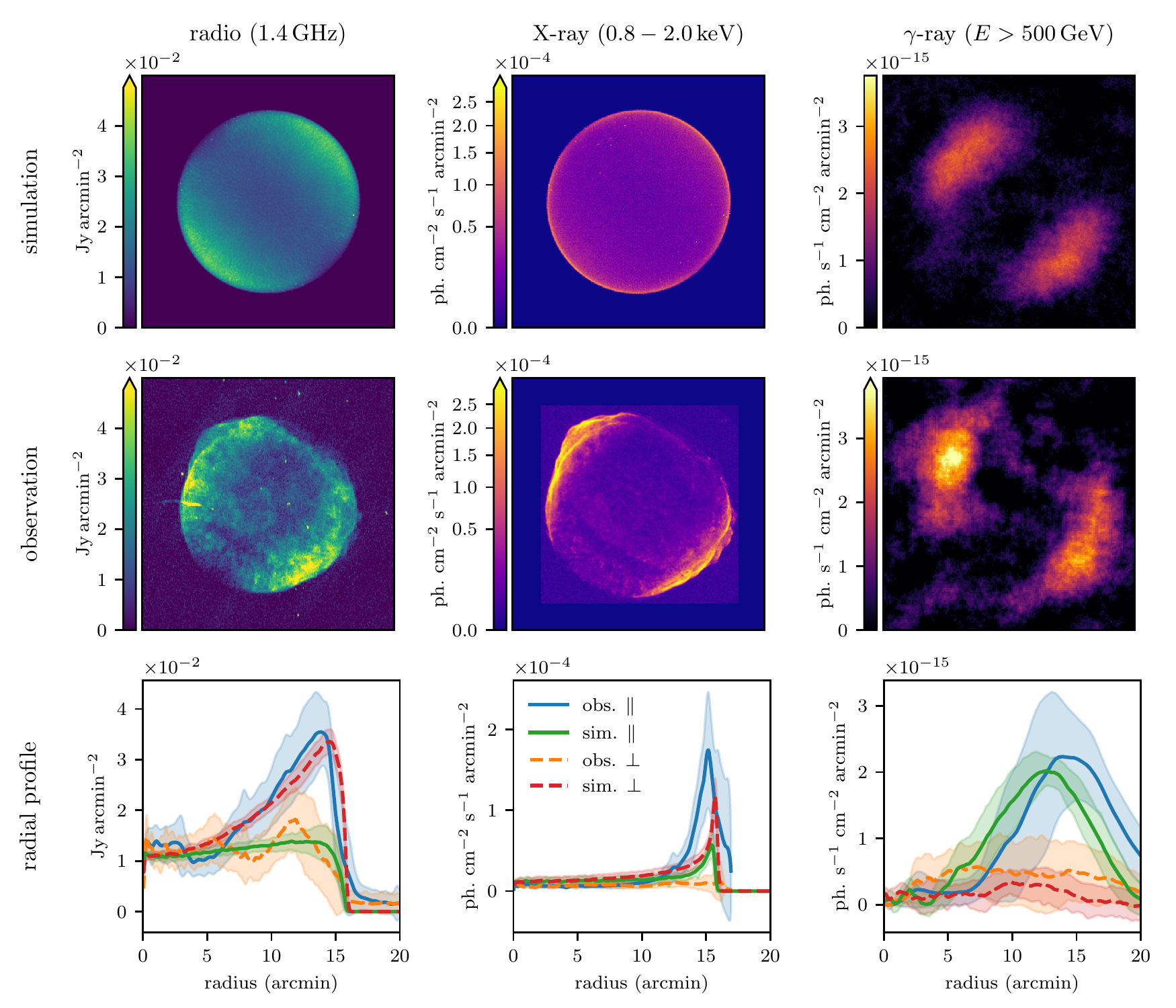}
	\caption{Morphological comparison of our simulation with a {\em constant, i.e., obliquity independent, acceleration}; details as in Figure~\ref{fig:Morphology_Parallel}. The leptonic synchrotron emission in the radio and X-rays does not match the observations in these bands \citep{Dyer2009, Cassam-Chenaie2008}. The simulated $\gamma$-ray map is dominated by the hadronic emission and in agreement with HESS observations \citep{Acero2010}. The maps have a side length of \SI{21}{pc} or \ang{;42.5;} at a distance of \SI{1660}{pc}.}
	\label{fig:Morphology_constant}
\end{figure*}

\begin{figure*}
	\includegraphics[width=\textwidth]{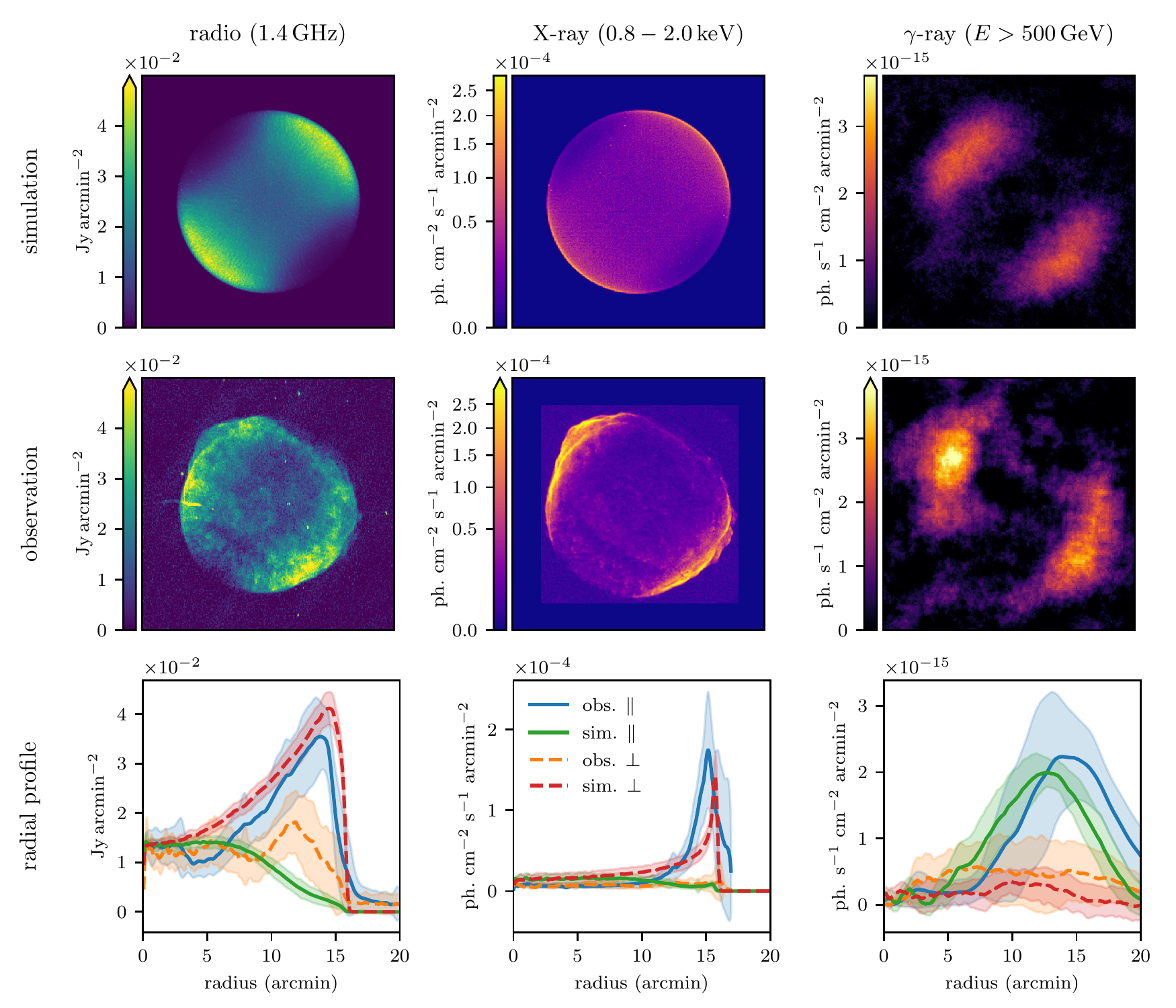}
	\caption{Morphology of the simulations with preferred {\em quasi-perpendicular acceleration} efficiency of CR electrons; details as in Figure~\ref{fig:Morphology_Parallel}. The leptonic synchrotron emission in the radio and X-rays does not match the observations in these bands \citep{Dyer2009, Cassam-Chenaie2008}. The simulated $\gamma$-ray map is dominated by the hadronic emission and in agreement with HESS observations \citep{Acero2010}. The maps have a side length of \SI{21}{pc} or \ang{;42.5;} at a distance of \SI{1660}{pc}.}
	\label{fig:Morphology_Perpendicular}	
\end{figure*}

In the previous section, we have presented a simulation with preferred quasi-parallel electron acceleration which matches the multi-frequency spectrum and the morphology of SN~1006. Because there is still an ongoing debate whether CR electrons can be efficiently accelerated in quasi-perpendicular or quasi-parallel configurations, we show that alternative acceleration scenarios are not able to reproduce the observed morphology. In the following, we critically compare our quasi-parallel acceleration model to simulations with constant, i.e., obliquity independent, and preferred quasi-perpendicular acceleration of CR electrons. However, in all cases, CR protons are accelerated in quasi-parallel configurations \citep{Caprioli2014}. For reference, the obliquity dependent acceleration efficiencies are presented in Figure~\ref{fig:Obliquity_Efficiency}.

\subsection{Quasi-parallel acceleration efficiency}

As explained in Section~\ref{sec:Leptonic_vs_Hadronic}, Figures~\ref{fig:BestFit_Spectrum} and \ref{fig:Morphology_Parallel}  show the total multi-frequency spectrum and the emission maps  of SN~1006 and demonstrate that overall this model provides a very good quantitative match to the observations while there are differences in detail. We expect that the inclusion of more realism in the simulations will also model these small-scale feature. In particular, including density fluctuations and small-scale interstellar turbulence so that the interaction with the shock produces a turbulent dynamo and magnetic field fluctuations may produce the observed patchy radio morphology and ripples in the shock surface. This could then explain the appearance of several shocks in projection in the X-ray surface brightness map. The same effect may then also slightly reduce the IC flux and improve the fit in the \Fermi band. Finally the asymmetry of the elongated $\gamma$-ray emitting regions being brighter in the North and dimmer in the South could originate from a large-scale gradient that boosts the hadronic pion-decay flux relative to the Southern counterpart \citep{Pais2020b}.

\subsection{Constant acceleration efficiency}

Figure~\ref{fig:Morphology_constant} shows the non-thermal emission maps of the
simulation with constant acceleration efficiency (top row), observations (mid
row), and radial profiles (bottom row) in the radio (\SI{1.4}{GHz}), X-ray
(\SIrange{0.8}{2}{keV}), and $\gamma$-ray band ($E > \SI{500}{GeV}$). The
simulation with constant acceleration efficiency uses the same parameters as
before with one exception: to fit the multi-frequency spectrum to the radio
data points, we need to adopt an electron acceleration efficiency of
$\zeta_\mathrm{e,max} = \num{2e-4}$.
It is apparent that the simulated radio and X-ray surface brightness maps have
bright regions in the SE and NW which do not match those of the
observations. Radial profiles illustrate this mismatch. On the contrary, the simulated
$\gamma$-ray surface brightness map is in agreement with observation as the
emission is dominated by CR protons which are accelerated at quasi-parallel
configurations. A rotation of the magnetic field by $90^\circ$ in the plane of
sky cannot resolve the mismatch in radio and X-ray as it would lead to a mismatch
in the $\gamma$-ray maps.

\subsection{Quasi-perpendicular acceleration efficiency}

Figure~\ref{fig:Morphology_Perpendicular} shows the non-thermal emission maps of the simulation with quasi-perpendicular acceleration (top row), observations (mid row) and radial profiles (bottom row). The
simulation with quasi-perpendicular acceleration efficiency uses the same parameters as
before with one exception: to fit the multi-frequency spectrum to the radio
data points, we need to adopt an electron acceleration efficiency of
$\zeta_\mathrm{e,max} = \num{2e-4}$. This acceleration scenario again leads to bright radio and X-ray regions in the SE and NW which are in disagreement with observations. However, there is an agreement for the $\gamma$-ray maps which are dominated by hadronic emission.

These two alternative obliquity dependencies for CR electron acceleration, i.e. constant and quasi-perpendicular, cannot reproduce the observed morphologies of SN~1006. This favours the preferred quasi-parallel acceleration of CR electrons.

\section{Amplification and damping of magnetic fields}
\label{sec:damping}
\begin{figure}
	\includegraphics[width=\columnwidth]{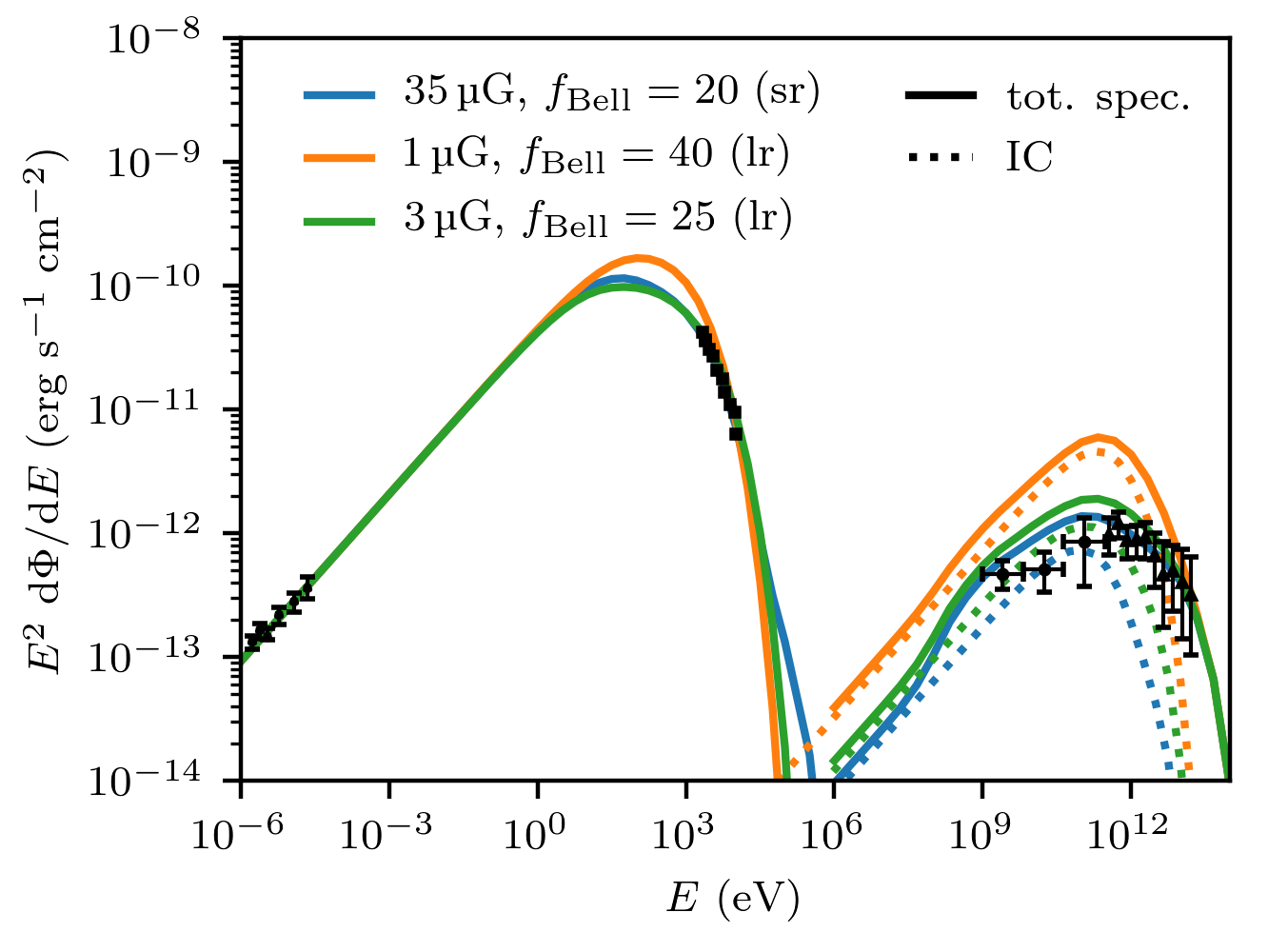}
	\caption{Multi-frequency spectra of models with short range (sr) and long range (lr) Bell-amplified magnetic fields. The spectrum in the $\gamma$-ray regime is a sum of IC emission (dotted) and hadronic $\gamma$-rays (not shown). The blue lines represent our best-fit model with volume-filling field (as result of a turbulent dynamo) and short range Bell-amplified fields. Orange and green lines represent models with typical ISM magnetic fields with long-range Bell-amplified magnetic fields but without a volume-filling amplification by a turbulent dynamo.}
	\label{fig:Spectra_Turbulent_vs_Bell}
\end{figure}
\begin{figure*}
	\includegraphics[width=\textwidth]{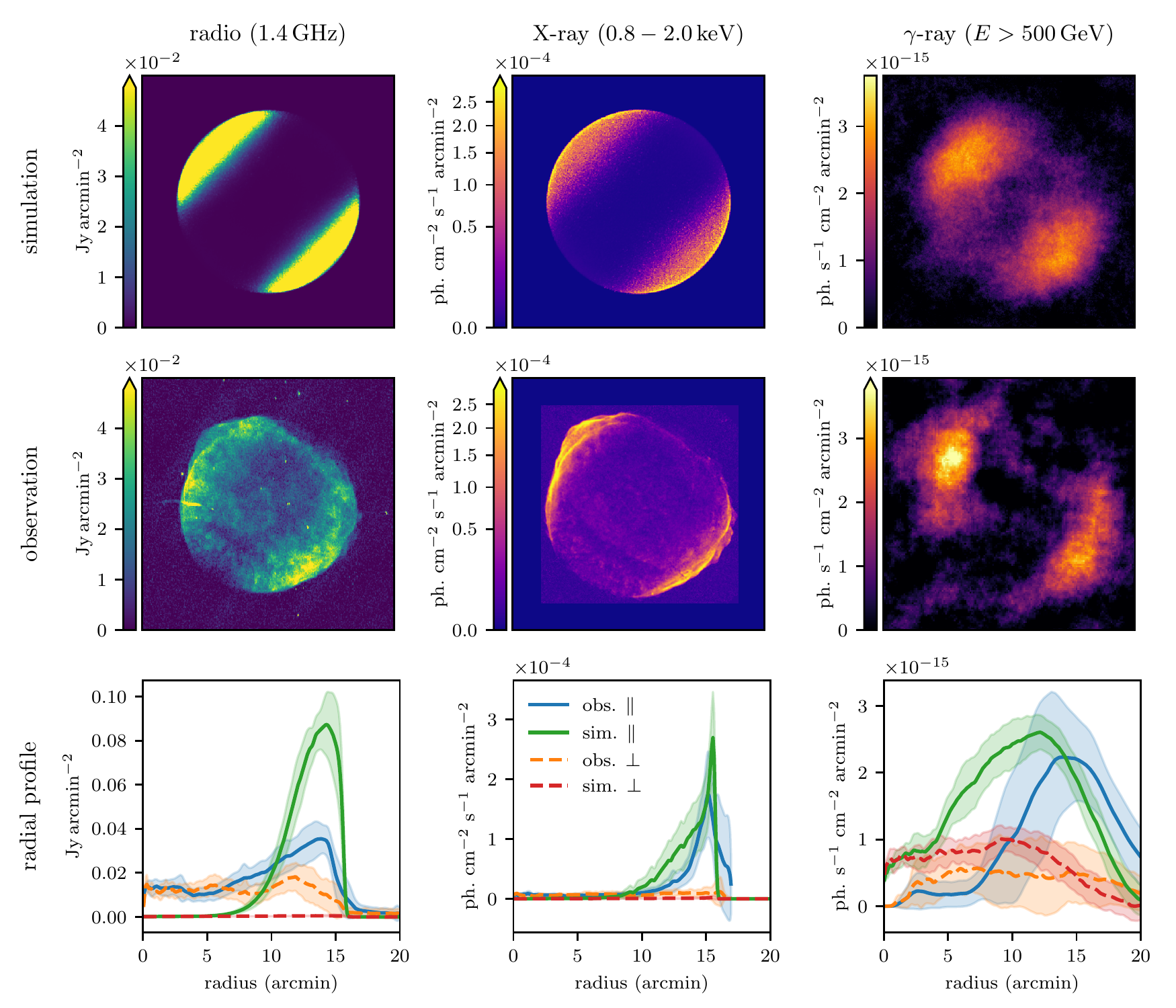}
	\caption{Morphology of simulations with a long-range Bell-amplified magnetic field that decays in proportion to the adiabatic expansion of the gas, $B= B_\mathrm{amp} n/n_\mathrm{shock}$ (because of its transverse polarization). The radio, X-ray, and $\gamma$-ray morphologies are in disagreement with observations. The maps have a side length of \SI{21}{pc} or \ang{;42.5;} at a distance of \SI{1660}{pc}.}
	\label{fig:Morphology_Decay}
\end{figure*}

\begin{figure*}
	\includegraphics[width=\textwidth]{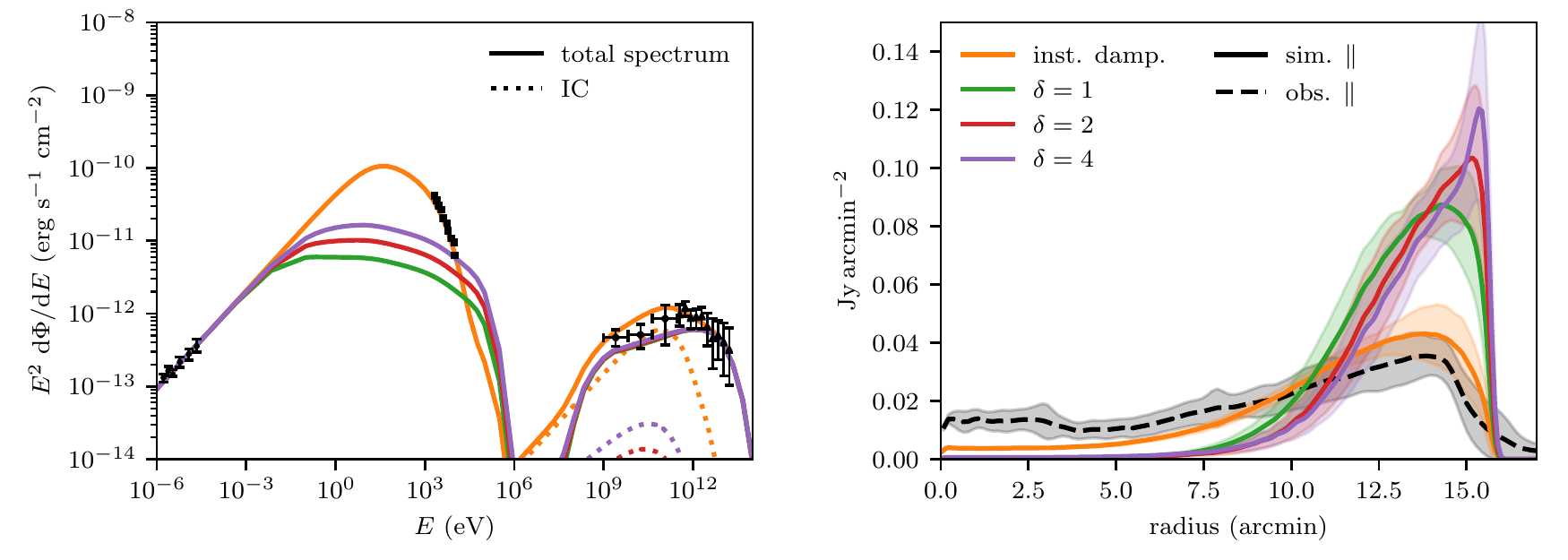}
	\caption{Multi-frequency spectra (left) and radial profiles of radio surface brightness maps (right) for different decay models of the Bell-amplified magnetic field via plasma effects. The model of instantaneous damping (orange solid) is our best-fit model and is in agreement with observations. Other models assume $B=(n/n_\mathrm{shock})^\delta B_\mathrm{amp}$.}
	\label{fig:BellDecay}
\end{figure*}

In this section, we demonstrate the case for volume-filling strong magnetic fields, potentially amplified by a turbulent dynamo, in order not to overproduce the $\gamma$-ray data points. In addition to these volume-filling magnetic fields in the post shock region, we model the amplification of magnetic fields via the non-resonant hybrid instability driven by CR protons in the upstream region close to the shock. These amplified magnetic fields decay due to strong ion-neutral collisional damping. In this section, we explain the influence of these fields and draw phenomenological conclusions on their damping length scale.

\subsection{Turbulent magnetic amplification}

As shown in Section \ref{sec:Leptonic_vs_Hadronic}, we obtain good agreement of the simulated multi-frequency spectrum with observations if there is a volume-filling amplified magnetic field (turbulently amplified fields) with an equivalent strength of $B=\SI{35}{\micro G}$ and if the field is additionally amplified on a short range by a factor of 20 directly at the shock via the non-resonant hybrid instability (Bell-amplified fields). In order to distinguish between the observational signatures of the different amplification processes, we first show simulations without any (volume-filling) turbulently amplified magnetic fields and present simulations in which the  Bell-amplified magnetic fields persist on a long range, i.e. they decay adiabatically with $B = (n / n_\mathrm{s}) B_\mathrm{amp}$ because the amplified Alfv\'en wave field is purely transverse to background magnetic field (and as such to the shock normal for parallel shocks). Here, $n$ denotes the current number density, $n_\mathrm{s}$ is the number density at the shock front, $B_\mathrm{amp}$ denotes the Bell-amplified magnetic field at the shock.

Figure~\ref{fig:Spectra_Turbulent_vs_Bell} compares the multi-frequency spectra of our best-fit (blue lines) to models with long-range Bell-amplified magnetic fields (orange and green lines). The models with long-range magnetic fields include typical values of the large-scale ISM magnetic field. A long-range Bell amplification by a factor $f_\mathrm{Bell} = 40$ together with an ISM field of \SI{1}{\micro G} (orange lines) leads to an overproduction of $\gamma$-ray emission as a large CR electron acceleration efficiency of $\zeta_{\mathrm{e}, \mathrm{max}} = \num{2.2e-3}$ is required in order to reproduce observational radio data. A larger ISM field of \SI{3}{\micro G} together with long-range Bell-amplified fields by a factor  a factor $f_\mathrm{Bell} = 25$ has a lower $\gamma$-ray emission that is close to observational $\gamma$-ray data ($\zeta_{\mathrm{e}, \mathrm{max}} = \num{8e-4}$).

Although this model with long-range Bell-amplified magnetic fields reproduces the observed multi-frequency spectrum fairly well it clearly fails to reproduce the observed morphology which is shown in Figure~\ref{fig:Morphology_Decay}. As before, we show simulations (top row), observations (middle row), and radial profiles (bottom row) for the radio (left column), X-ray (central column), and $\gamma$-ray band (right column).

The simulated radio map (top left panel of Fig.~\ref{fig:Morphology_Decay}) has two polar caps which are brighter and more confined in comparison to our best-fit simulation with short-range Bell-amplified magnetic fields in Figure~\ref{fig:Morphology_Parallel}. This is due to the sustained amplified magnetic fields which dramatically increase the synchrotron luminosity at radio frequencies. Although we chose an CR electron maximal acceleration efficiency to be $\zeta_\mathrm{e, max} = \num{8e-4}$ such that the simulated spectrum fits the radio data, the radial profile (bottom left of Fig.~\ref{fig:Morphology_Decay}) shows a clear mismatch to the radio observations. This is a consequence of the fast synchrotron cooling of CR electrons in the strong magnetic fields so that the central regions of the simulated remnant are devoid of radio emission.

The simulated X-ray map (top centre panel of Fig.~\ref{fig:Morphology_Decay}) has a similar morphology in comparison to the simulated radio map. It shows bright polar caps that are significantly wider than the observed X-ray rims (mid centre panel). In addition, the simulated $\gamma$-ray map (right column) is in disagreement with observations and fill in the central parts of SN~1006 unlike the HESS observations. Leptonic $\gamma$-rays are contributing significantly to the emission at $E \approx \SI{500}{GeV}$ because the small volume of the radio-emitting regions require a larger CR electron acceleration efficiency in comparison to our best-fit model with short-range Bell-amplified fields.

We therefore conclude that a volume-filling magnetic field, potentially amplified by a turbulent dynamo, is necessary in order to reproduce the observed multi-frequency spectrum and morphology.

\subsection{Magnetic amplification via the Bell instability}

As shown before, we obtain good agreement of the simulated multi-frequency spectrum with observations if magnetic fields are amplified by a factor of 20 directly at the shock and decay immediately behind it. To determine the sensitivity of the non-thermal emission maps on the phenomenological model of the magnetic field decay, we present simulations in which the amplified magnetic fields decay significantly slower and adopt a scaling with the gas density according to
\begin{align}
B = \left(\frac{n}{n_\mathrm{s}} \right)^\delta B_\mathrm{amp},
\end{align}
where $\delta$ is the damping parameter.

Figure~\ref{fig:BellDecay} shows multi-frequency spectra (left) and radial profiles of the radio maps (right) for our best-fit model with instantaneous damping and three models with different decay parameters $\delta$. It is evident, that only the simulation with instantaneous damping of amplified magnetic fields shows good agreement with the observed multi-frequency spectrum and the radial profile.
Larger decay parameters lead to an increasing X-ray and $\gamma$-ray emissivity because CR electrons suffer strong cooling losses on shorter timescales. However, these simulations significantly deviate from observed profiles.

We therefore conclude that amplified magnetic fields driven by CR proton current at the shock have to decay on a very short length scale close to our discretized Voronoi cell size at the shock (corresponding to 100 gyroradii for TeV particles) and cannot be sustained for a long time in the post shock region.

\section{Parameter dependencies}
\label{sec:parameters}
The multi-frequency spectrum is influenced by several parameters. We first study the dependence of the spectra on different CR proton and electron spectral indices. Secondly, we present spectra for varying equivalent magnetic field strengths (possibly as a result of turbulent amplification), Bell amplification factors, maximum acceleration momentum. Finally, we study the influence of ambient photon fields and gas densities.

\subsection{Spectral index of CRs}
\begin{figure}
	\includegraphics[width=\columnwidth]{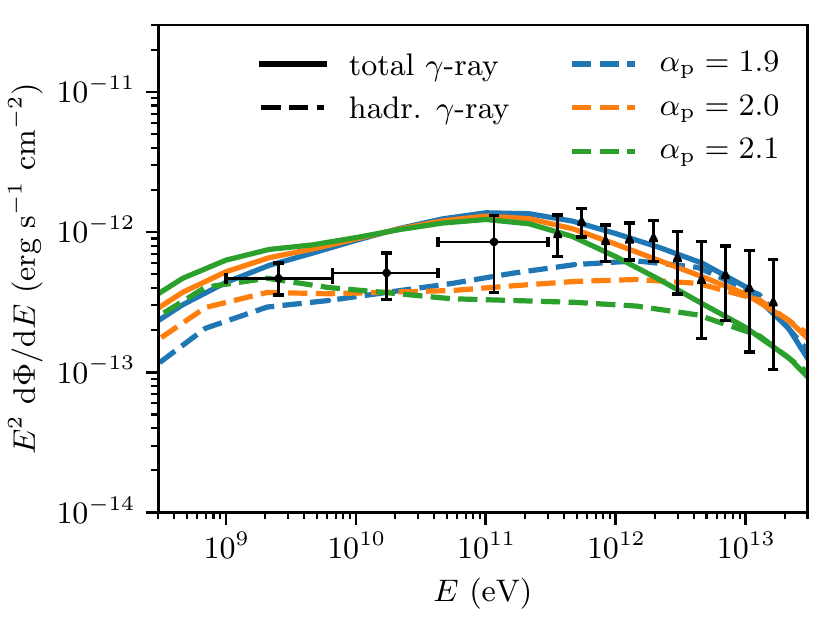}
	\caption{Total $\gamma$-ray spectra (solid lines) and hadronic $\gamma$-ray spectra (dashed lines) with different CR proton spectral indices $\alpha_\mathrm{p}$ are compared to \Fermi and HESS data. A best-fit to observational data is given for a CR proton spectral index of $\alpha_\mathrm{p} = 1.9$.}
	\label{fig:Spectrum_Proton_Variation}
\end{figure}

Figure~\ref{fig:Spectrum_Proton_Variation} shows the total (solid lines) and hadronic (dashed lines) high energy $\gamma$-ray spectra for CR proton spectral indices of $\alpha_\mathrm{p} = 1.9$, 2.0, and 2.1. We use a maximum CR proton momentum of $p_\mathrm{max} = \num{2.1e5}$ for $\alpha_\mathrm{p} = 1.9$ and $p_\mathrm{max} = \num{4.2e5}$ for the latter two indices (see equation~\ref{eq:ProtonSpectrum}).
 We adopt our best-fit leptonic spectrum with a CR electron spectral index of $\alpha_\mathrm{e} = 2.1$  (see Figure~\ref{fig:BestFit_Spectrum}) to calculate the total spectrum. The hadronic spectrum for $\alpha_\mathrm{p} = 1.9$ agrees best with the observed data, which is especially visible for the first \Fermi data point (at $\approx \SI{3}{GeV}$) and for the HESS data points above \SI{1}{TeV}. A steeper proton spectral index leads to an overestimate of the $\gamma$-ray flux at GeV energies and at the same time to an underestimate at TeV energies. Hence, we use the best fitting value of $\alpha_\mathrm{p} = 1.9$ for further analysis of the spectrum.

\begin{figure*}
	\includegraphics[width=\textwidth]{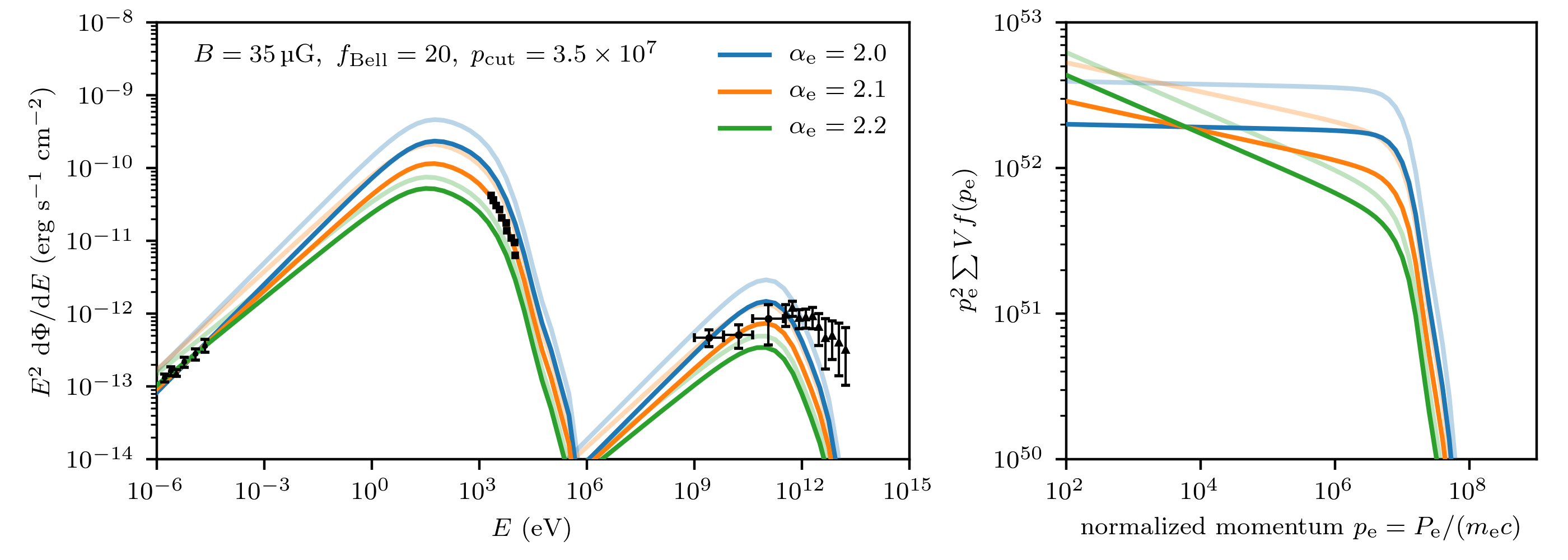}
	\caption{Multi-frequency spectra (left) and CR electron spectra (right) for different CR electron spectral indices $\alpha_\mathrm{e}$. The multi-frequency spectra show only the leptonic emission and use only CMB photons for the IC calculation. Semi-transparent lines show the results using a maximal electron acceleration efficiency of $\zeta_\mathrm{e, max} =10^{-3}$ and opaque lines show spectra where $\zeta_\mathrm{e, max}$ is chosen such that they are in agreement with observational radio data.}
	\label{fig:Spectrum_Slope_Variation}
\end{figure*}

We show the influence of the CR electron spectral index $\alpha_\mathrm{e}$ in Figure~\ref{fig:Spectrum_Slope_Variation}. Other parameters such as the equivalent magnetic field of $B=\SI{35}{\micro G}$, the amplification factor of 20, and the maximum acceleration momentum of $p_\mathrm{cut}=3.5\times10^{7}$ remain fixed.
The left-hand panel of Figure~\ref{fig:Spectrum_Slope_Variation} shows the multi-frequency spectrum. For clarity, we show only the IC spectrum on CMB photons in the $\gamma$-ray range.  The panel on the right hand side shows the total volume-weighted CR electron spectrum. Semi-transparent lines show the result for a fixed CR electron acceleration efficiency of $\zeta_\mathrm{e, max} = 10^{-3}$ in both panels. Opaque lines show the same model, however with a renormalised CR electron acceleration efficiency such that the  spectral radio data is fit. Acceleration efficiencies $\zeta_{\mathrm{e}, \mathrm{max}}$ of renormalised spectra are $\num{5.1e-3}$ for $\alpha_\mathrm{e} = 2.0$, $\num{5.4e-3}$ for $\alpha_\mathrm{e} = 2.1$, and $\num{7.0e-3}$ for $\alpha_\mathrm{e} = 2.2$.

In the following discussion, we refer to opaque lines with floating acceleration efficiency thus ensuring a match to radio data. A CR electron spectral index of $\alpha_\mathrm{e} = 2.0$, which is the test-particle limit of DSA theory, leads to an overestimate of the $\gamma$-ray flux at energies of 10 to \SI{100}{GeV}. Therefore a larger spectral index $\alpha_\mathrm{e} > 2.0$ is necessary in order to produce an agreement with $\gamma$-ray data. However, a spectral index of $\alpha_\mathrm{e} = 2.2$ leads to an underestimate of X-ray data which cannot be compensated by having a larger acceleration momentum $p_\mathrm{cut}$ because of the different spectral shape. Hence, a spectral index of $\alpha_\mathrm{e} = 2.1$ is our best fit which produces results compatible with X-ray and $\gamma$-ray data.

We have shown the influence of the CR electron and proton spectral index on the spectrum and that a good agreement with observations is obtained with the spectral indices $\alpha_\mathrm{e} = 2.1$ for electrons and $\alpha_\mathrm{p} = 1.9$ for protons. We use these two best-fit values throughout the rest of our parameter study.

There are several mechanisms which potentially lead to spectral indices that are different from the canonical value of $\alpha=2$ in the test-particle limit of diffusive shock acceleration (DSA) and to slight deviations of the electron and proton spectral indices.
First, accelerated CRs provide a pressure component in addition to thermal pressure that changes the shock structure which is referred to as non-linear DSA \citep[e.g.][]{Eichler1979, Bell1987, Amato2005, Reynolds2008}. The incoming flow is gradually decelerated in a dynamical precursor that is generated by CRs diffusing ahead of the shock. A thermal subshock remains but the overall compression ratio from far upstream to downstream is increased. Low energetic particles, i.e., CR electrons and CR protons with small momenta, obtain a softer spectral index by experiencing a weaker shock. Particles with greater particle energies have a longer mean free path between scattering events and are able to experience the larger compression ratio.
Second, the spectral index of accelerated particles in DSA depends on the escape probability \citep{Bell1978a} which can be estimated with the diffusive spatial transport \citep{Kirk1996, Lazarian2014}. It is found that the spectral index follows the relation $\alpha = 3r/(r-1)\times[1 + (1-\beta)/r] - 2$ where $r$ is the shock compression ratio and $\beta$ determines the diffusive transport via the relation $\langle \Delta x^2 \rangle \propto t^\beta$. Standard diffusion with $\beta=1$ gives $\alpha=2$ but anomalous transport in turbulent fields yields different diffusion schemes and different spectral indices \citep{Duffy1995, Lazarian2014}.
Third, the spectrum of accelerated particles at SNRs can be steepened by geometric and time-dependent processes \citep{Malkov2019b} or non-local processes associated with changes of the magnetic field orientation along the shock front \citep{Hanusch2019}.
Fourth, accelerated CRs self generate electric and magnetic fields which lowers the energy of the CRs and thereby modifies the spectrum \citep{Zirakashvili2015, Osipov2019b, Bell2019}, potentially in a way that is different for electrons and protons \citep{Bell2019}.
Fifth, the inclusion of higher-order anisotropies of the CR spectrum near to shock shows that the spectral index changes as a function of magnetic obliquity and shock velocity \citep{Bell2011, Takamoto2015}.

\subsection{Magnetic amplification and maximum momentum}

\begin{figure*}
	\includegraphics[width=\textwidth]{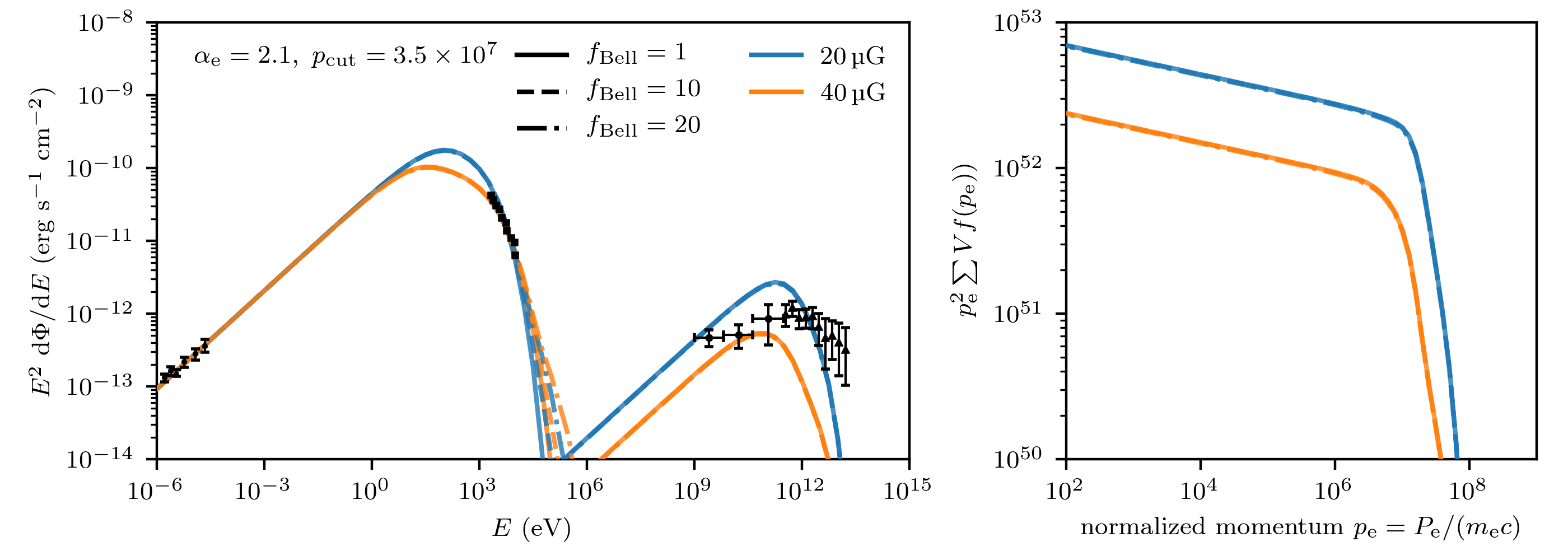}
	\includegraphics[width=\textwidth]{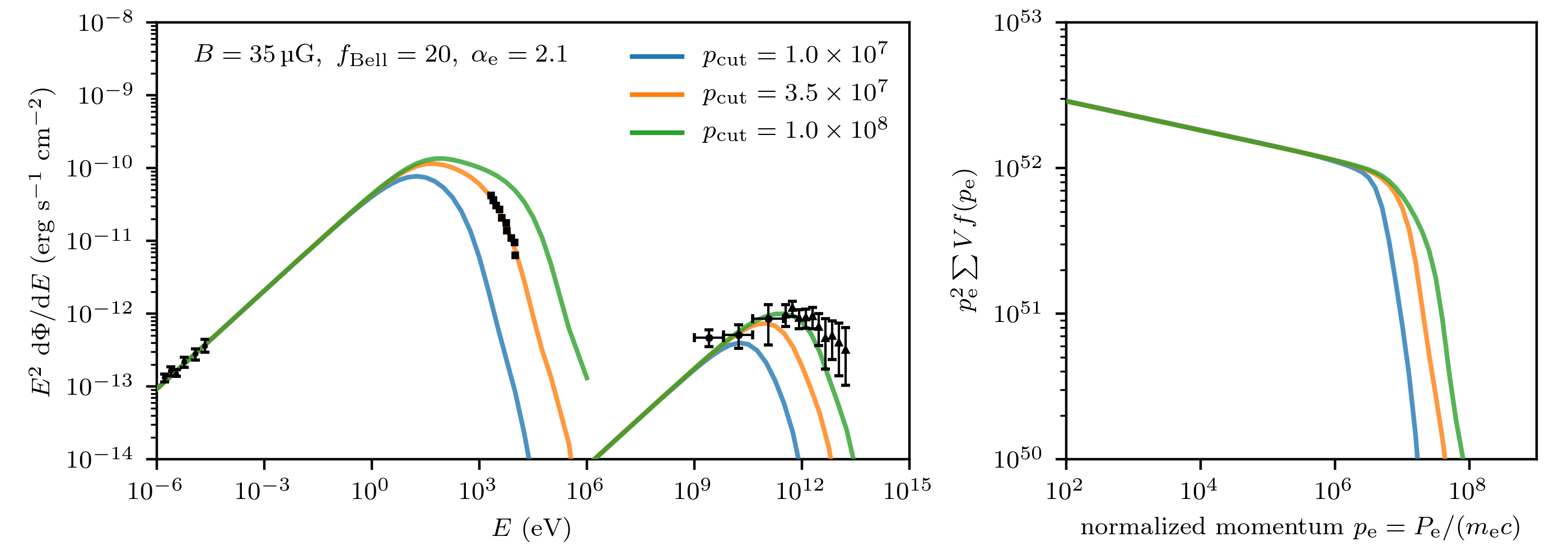}
	\caption{Multi-frequency spectra (left) and CR electron spectra (right) for different equivalent magnetic fields and Bell amplification factors (top row) and for different maximal acceleration momentum $p_\mathrm{cut}$ of CR electrons (bottom row). The acceleration efficiency is chosen such that the spectra fit observed radio data.}
	\label{fig:B_Field_Variation}
\end{figure*}

\begin{figure*}
	\includegraphics[width=\textwidth]{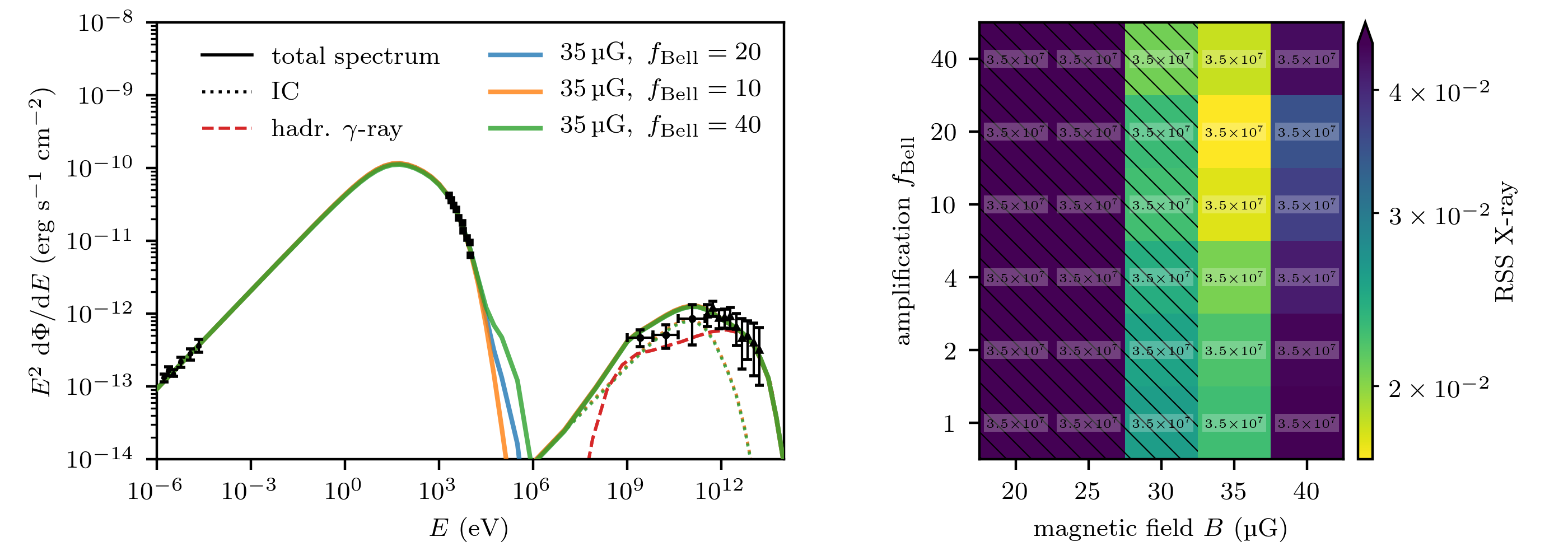}
	\caption{Best fit multi-frequency spectra (left) and parameter space (right) that is spanned by an equivalent magnetic field ($x$-axis), Bell amplification factor ($y$-axis), and maximum CR electron momentum (insets). The colour bar on the right hand side indicates the quality of fit to X-ray data in terms of the residual sum of squares (RSS). The RSS values are calculated in logarithmic space. Hatched parameter combinations indicate that simulations produce high $\gamma$-ray emission exceeding 2.5$\sigma$ uncertainties of observational data. This motivates our choice of \SI{35}{\micro G} and a Bell amplification factor 10 for our best-fit model.}
	\label{fig:BestFit}
\end{figure*}

We move on to study the influence of magnetic amplification and maximum CR electron momentum on the spectrum. In the following, we always refer to spectra that are obtained with a free floating CR electron acceleration efficiency $\zeta_\mathrm{e, max}$ such that a fit to spectral radio data is obtained.

The top row of Figure~\ref{fig:B_Field_Variation} shows how different equivalent magnetic fields and Bell amplification factors $f_\mathrm{Bell}$ shape the multi-frequency spectrum (left) and which CR electron spectrum (right) is necessary to fit the radio data.  For clarity, the multi-frequency spectrum only contains the leptonic spectra together with the IC emission on CMB photons.
A strong magnetic field leads to fast cooling CR electrons such that the synchrotron spectrum is reduced at photon energies $E \gtrsim \SI{10}{eV}$ while extending its tail to a slightly larger energy as can be seen in the left-hand panel. The orange lines representing simulations with a \SI{40}{\micro G} field deviate at lower energies from the synchrotron power law in comparison to the blue lines representing simulations with a $\SI{20}{\micro G}$ field. The Bell amplification factor has only a minor influence on the synchrotron spectrum because these Bell-amplified fields are constrained to a small volume at the shock front. The acceleration efficiencies $\zeta_{\mathrm{e}, \mathrm{max}}$ are \num{1.3e-2} for the \SI{20}{\micro G} equivalent field and \num{4.3e-3} for \SI{40}{\micro G}.

The top right panel of Figure~\ref{fig:B_Field_Variation} shows that the CR electron spectrum of the \SI{40}{\micro G} simulation (orange line) is lower than that of the \SI{20}{\micro G} (blue line) because a larger magnetic field requires a lower CR electron acceleration efficiency in order to fit observed radio data. This results in a lower CR electron spectrum which implies a lower IC emissivity as can be seen in the left-hand panel. Consequently, low magnetic fields with larger CR electron acceleration efficiencies are excluded because they overestimate the high-energy $\gamma$-ray spectrum.

The bottom row of Figure~\ref{fig:B_Field_Variation} shows the influence of the maximum acceleration momentum $p_\mathrm{cut}$ of CR electrons. The panel on the left-hand side shows the multi-frequency spectrum while the right-hand side shows the CR electron spectrum. Note that while we fix $p_\mathrm{cut}$ for a given simulation, the effective spectral cutoff of our Lagrangian particles is dynamically evolving due to adiabatic processes and cooling losses so that the final cutoff of the total spectrum is a superposition of all individually transported spectral cutoffs. It is apparent that the maximum acceleration momentum is important for obtaining an agreement with spectral X-ray data. A too small maximum acceleration momentum underestimates the synchrotron spectrum at X-ray energies whereas a too large value leads to an overestimate. The cooling of the CR electron spectrum due to synchrotron and IC losses cannot compensate a too large maximum acceleration momentum because it leads to flattening of the synchrotron spectrum rather than a cutoff as suggest by the data.

We have shown, that the value of the turbulently amplified magnetic field, the Bell amplification factor, and the maximum acceleration momentum are essential for obtaining a multi-frequency spectrum that is in agreement with observations. We now extend our study to a larger parameter space of these values considering now hadronic $\gamma$-rays as well. Figure~\ref{fig:BestFit} shows the result of this study for the best-fit values of the spectral indices $\alpha_\mathrm{p} = 1.9$ for CR protons and $\alpha_\mathrm{e} = 2.1$ for electrons. The panel on the left-hand side shows the three best model which are in agreement with observations. Solid lines represent the total spectrum while dotted and dashed lines show the leptonic and hadronic $\gamma$-ray spectrum, respectively.

The right-hand panel of Figure~\ref{fig:BestFit} shows the residual sum of squares (RSS) at X-ray energies in the parameter space of magnetic field and amplification factor indicated by the colours from yellow (good fit) to purple (bad fit). The RSS values are calculated with the logarithmic spectral values as they span an order of magnitude.
For each combination of equivalent magnetic field and Bell amplification factor, we report the best-fit value of the maximum acceleration momentum of CR electrons. Hatched cells represent parameter combinations that overproduce the spectrum at $\gamma$-ray energies, i.e., a total spectrum exceeding $2.5\sigma$ of at least one \Fermi or HESS data point.

The parameter combination of an equivalent magnetic field of $B=\SI{35}{\micro G}$, a Bell amplification factor of 20, and a maximum acceleration momentum of $p_\mathrm{cut}= 3.5 \times 10^7$ produces the best agreement with X-ray data while being compatible with $\gamma$-ray data. By construction, they also fit radio data. We note that there is some degeneracy between these values as well as other parameters, e.g., density, explosion energy and CR spectral index. Hence, slightly different combinations might result in similar agreement with observational data. However, certain ranges of magnetic fields strengths can be excluded because they either overestimate $\gamma$-ray data, e.g., combinations of a low magnetic fields and a large acceleration efficiency (see Figure~\ref{fig:B_Field_Variation}), or they underestimate X-ray data due to fast cooling of CR electrons in strong magnetic fields. We conclude that volume-filling magnetic fields of $B\approx\SI{35}{\micro G}$ (possibly amplified through a turbulent small-scale dynamo) produce a good agreement with observations.

\begin{figure*}
	\includegraphics[width=\textwidth]{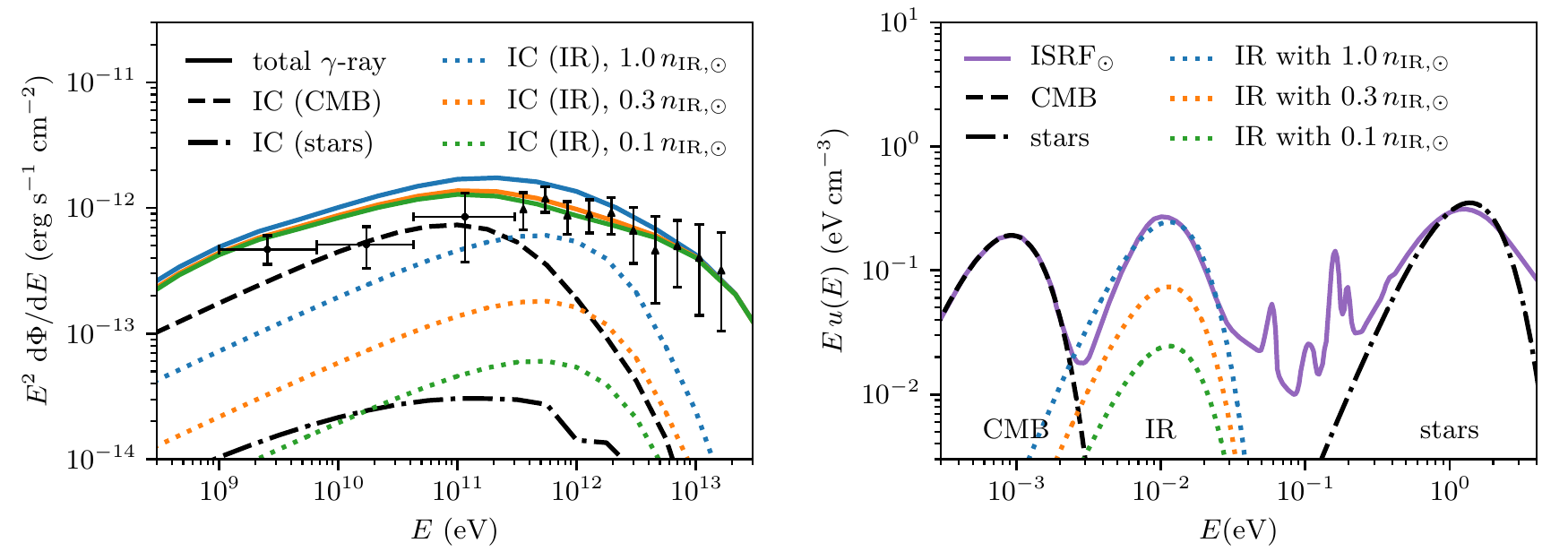}
	\caption{High energy $\gamma$-ray spectrum (left) and interstellar radiation fields (ISRF) (right). On the left hand side, the solid lines show the total $\gamma$-ray spectrum as sum of three IC components: CMB (dashed), IR (dotted), and star light (dash-dotted). Blue, orange, and green lines represent variable IR components where $n_{\mathrm{IR},\odot}$ is the density of infrared photons at the solar radius. On the right hand side, we show the ISRF at solar radius \citep{Porter2005,Porter2008} and our three-component ISRF model used for IC calculation.}
	\label{fig:RadiationFields}
\end{figure*}

\subsection{Ambient photon field and density}

We have discussed how the magnetic field indirectly influences the IC spectrum via the CR electron acceleration efficiency. We now discuss the direct influence of radiation fields on the IC spectrum. Figure~\ref{fig:RadiationFields} shows the high energy $\gamma$-ray spectrum in the left-hand panel for three different photons fields which are shown in the right-panel together with the interstellar radiation field models at different locations in the Milky Way.
The blue lines represent the spectrum that is obtained by fitting three black body spectra to the radiation field at the solar radius. Orange and green lines represent variations where $n_\mathrm{IR}$ is given by $0.3\, n_{\mathrm{IR}, \odot}$ and $0.1 \times n_{\mathrm{IR}, \odot}$, respectively. It is apparent that an infrared field similar to that of the solar radius leads to large total $\gamma$-ray spectrum (blue lines, left) exceeding $\gamma$-ray data from \Fermi and HESS. The contribution of the IC spectrum produced by interaction of CR electrons with starlight photons is negligible as it is suppressed due to the Klein-Nishina effect. Lower infrared fields with $n_\mathrm{IR} \lesssim 0.3 \times n_{\mathrm{IR}, \odot}$  (orange, green) lead to a good agreement of the the total $\gamma$-ray spectrum with observations.

\begin{figure*}
	\includegraphics[width=\textwidth]{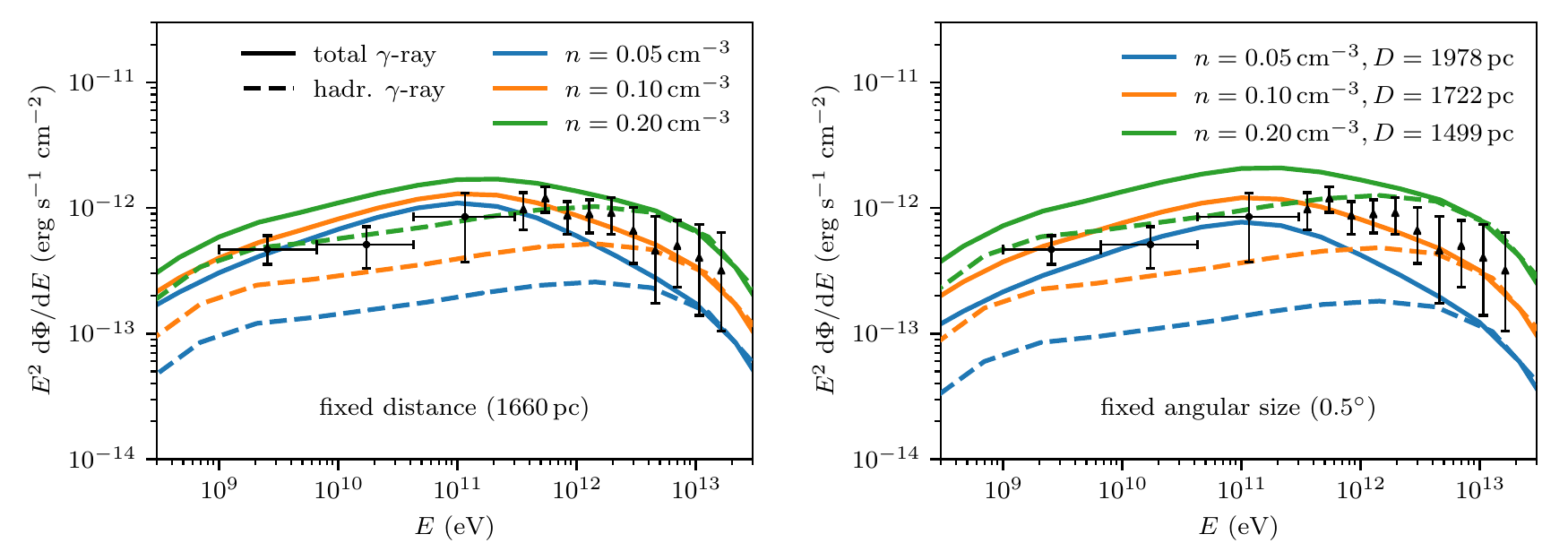}
	\caption{High energy $\gamma$-ray spectrum for varying ambient gas densities at a fixed distance of \SI{1660}{pc} (left) and at a fixed angular size of $0.5^\circ$ (right).}
	\label{fig:Density_Variation}
\end{figure*}

Finally, we explain the influence of ambient gas density onto the gamma-ray spectrum. Figure~\ref{fig:Density_Variation} shows the total $\gamma$-ray spectrum (solid lines) together with the hadronic $\gamma$-ray spectrum (dashed lines) for a fixed distance of \SI{1660}{pc} to the remnant (left) and for a fixed angular size of $0.5^\circ$ (right). The panel on the left-hand side shows the direct effect of a reduced target proton density for hadronic $\gamma$-ray production because it is directly proportional to the ambient gas density. Hence, the simulation with a low number density of $n = \SI{0.05}{cm^{-3}}$ (blue lines) underestimates the $\gamma$-ray flux for $E \gtrsim \SI{1}{TeV}$ whereas the simulation with a high number density of $n = \SI{0.2}{cm^{-3}}$ (green lines) leads to an overestimate.

However, we cannot choose the density as a free parameter and must also take into account the size of the remnant which is larger for lower densities. The radius of the remnant in the adiabatic phase evolves with
\begin{align}
R \propto \left(\frac{E_\mathrm{SN} t^2}{n} \right)^{1/5}
\end{align}
according to the Sedov--Taylor solution where $E_\mathrm{SN}$ is the SNR explosion energy. If we fix the angular size of the SNR to the observed solid angle, the distance $D$ has to scale in proportion with the radius. Consequently, the spectrum is influenced and scales according to
\begin{align}
\frac{\mathrm{d}\Phi}{\mathrm{d}E} \propto D^{-2} \propto n^{2/5}.
\end{align}
This is shown in the right-hand panel of Figure~\ref{fig:Density_Variation}, where a low (high) density leads to an even stronger underestimate (overestimate) of the $\gamma$-ray spectrum. The resulting distances are given in the plot.

Our best fit value $n=\SI{0.12}{\per\cubic\centi\meter}$ for the ambient density is well within the statistical and systematic uncertainties of the observations. Analysis of the SE rim with \textit{XMM-Newton} yields post-shock densities from $n_\mathrm{post} = 1.44^{+0.10}_{-0.11}$ to $1.99^{+0.4}_{-0.17}\,\si{cm^{-3}}$ with the larger values arising due to CR shock modification and a pre-shock density of $n_\mathrm{pre} \approx \SI{0.04}{cm^{-3}}$ \citep{Miceli2012} which is in agreement with $n_\mathrm{pre} = \SI{0.05}{cm^{-3}}$ obtained by \citet{Acero2007}. Comparable results are inferred from \textit{Chandra} data which show that the pre-shock density is $n_\mathrm{pre} = 0.045^{+0.049}_{-0.020}\,\si{cm^{-3}}$ \citep{Winkler2014}. Higher densities are found in the western part of the SN~1006 where the remnant interacts with an atomic cloud in the SW and an H$\alpha$-bright cloud in the NW. In the NW, analysis of \textit{Spitzer} data gives a post-shock density of $n_\mathrm{post} = 1.4\pm\SI{0.5}{cm^{-3}}$ which yields a pre-shock density of $n_\mathrm{pre} = 0.35 \pm \SI{0.125}{\cm^{-3}}$ using a standard shock compression ratio of four \citep{Winkler2013}. In the SW, analysis of of X-ray data yields \SIrange{0.3}{0.5}{cm^{-3}} for the pre-shock density \citep{Miceli2014, Miceli2016}.

\section{Discussion and conclusion}
\label{sec:discussion}
We have performed 3D MHD simulations of the remnant of SN~1006 with CR proton and electron physics which includes the spatial and temporal evolution of the CR electron spectrum. We account for leptonic emission processes, i.e., synchrotron and IC emission, and hadronic $\gamma$-ray emission, and present multi-frequency spectra and non-thermal emission maps in the radio, X-ray, and $\gamma$ rays. We model the magnetic obliquity dependent CR proton acceleration following results of hybrid particle-in-cell simulations of \citealt{Caprioli2014}). In addition, we study different models of obliquity dependent CR electron acceleration (some of which are also inspired by recent particle-in-cell simulations) and explore the influence of various model parameters on the maps and non-thermal emission spectra.

Our main conclusions are summarised here.
\begin{itemize}
\item Because our simulations lack the dynamic rage to fully resolve a
  turbulent dynamo caused by small-scale density fluctuations in the
  interstellar medium, and our model of the CR physics precludes the excitation
  and growth of the non-resonant hybrid instability \citep{Bell2004}, we model
  these processes in form of a subgrid model. To this end, we evoke a turbulent
  dynamo (or a similar plasma process) behind the shock to generate a
  volume-filling magnetic field inside the SNR with values
  of $B=\SI{35}{\micro G}$ for perfectly parallel configurations up to
  $B=\SI{140}{\micro G}$ for perfectly perpendicular configurations. The
  magnetic field quickly increases with magnetic obliquity $\theta$ as shown in
  Figure~\ref{fig:Obliquity_Magnetic_Field} such that the magnetic field
  is \SI{76}{\micro G} at an obliquity angle of $\theta = 30^\circ$. Our
  simulations represent the case of a completely homogeneous initial magnetic
  field but small scale local fluctuations of the initial magnetic field lead to
  locally different obliquity angles and locally larger field
  strengths \citep{Pais2020}. However, the detailed modeling of a turbulent
  magnetic field component superposed on the dominating homogeneous field in the
  initial conditions is beyond the scope of this work.  Averaging over areas
  with different obliquity angles as a result of line-of-sight projection and
  small-scale magnetic turbulence can explain magnetic fields on the order
  of \SI{100}{\micro G} in quasi-parallel regions as inferred from the analysis
  of X-ray filaments in the NE of SN~1006 \citep{Morlino2010} and in the NE and
  SW limbs \citep{Ressler2014}.
  
\item In this best-fit model, we additionally account for the amplification
  of magnetic fields by a factor of about 20 due to Bell's instability (with the
  same obliquity dependence as we adopt for the CR proton acceleration
  efficiency) and assume that the SNR expands into a homogeneous medium on large
  scales with an average gas number density of $n=\SI{0.12}{cm^{-3}}$.

\item Leptonic and hadronic $\gamma$-ray emission are both important for
  explaining the observed $\gamma$-ray spectrum. In our model, hadronic
  pion-decay and leptonic emission (primarily from Compton-upscattering of CMB
  photons) are contributing to the emission at GeV $\gamma$-ray energies
  accessible to the \Fermi $\gamma$-ray space telescope approximately by equal
  parts. Within our adopted large parameter space, we find no solution with a
  smaller IC $\gamma$-ray component that simultaneously matches the
  multi-frequency spectrum and the non-thermal emission maps. However, hadronic
  emission is dominating at TeV energies that are observable by imaging air
  Cherenkov telescopes. We find, that the HESS $\gamma$-ray map at photon
  energies $E > \SI{500}{GeV}$ is thus dominated by hadronic pion-decay
  emission.

\item The model of preferentially {\em quasi-parallel shock acceleration of CR
  electrons} produces non-thermal emission maps and a multi-frequency spectrum
  that are in very good agreement with all observations.  In this model, the
  electron acceleration efficiency of radio-emitting GeV electrons at
  quasi-perpendicular shocks is suppressed at least by a factor ten.  The models
  of obliquity independent and {\em preferentially quasi-perpendicular shock
  acceleration} produce radio and X-ray maps that are in disagreement with
  observations. Because the simulated $\gamma$-ray map, which is dominated by
  hadronic emission, agrees with the observation, a rotation of the large-scale
  magnetic field by $90^\circ$ in the plane of sky cannot resolve this
  disagreement. Hence, this precludes extrapolation of current 1D plasma
  particle-in-cell simulations of particle acceleration at SNR shock conditions
  that favour preferentially quasi-perpendicular electron acceleration at
  shocks.

\item The low level of observed $\gamma$-ray flux requires a volume-filling
  strong magnetic field so that most of the electron energy is emitted via
  synchrotron emission. The preference of quasi-parallel acceleration of protons
  and electrons argues for efficient amplification of magnetic fields via Bell's
  instability (or a similar plasma process). We demonstrate that these
  Bell-amplified magnetic fields have to decay on short length scales of order
  100 gyroradii for TeV particles. Otherwise, CR electrons are subject to
  strong synchrotron losses which would lead to extended radial profiles of the
  radio and X-ray synchrotron emission at the shock that are in disagreement with
  observations. However, the exact value of the Bell amplification factor is
  only weakly constrained by the total spectrum because those amplified fields
  are confined to a small emission volume around the shock front.

\end{itemize}
Our work opens up a new avenue to study the physics of electron acceleration at shocks
and connects plasma physics at collisionless shocks to astrophysical scales of
SNRs in a novel and innovative manner.

\section*{Acknowledgements}
It is a pleasure to thank Volker Springel for the use of \textsc{arepo}. 
We warmly thank R\"udiger Pakmor and Damiano Caprioli for discussions and our anonymous referee for an insightful report that helped to improve the paper.
We acknowledge support by the European Research Council under ERC-CoG grant
CRAGSMAN-646955.

\section*{Data availability}
The data underlying this article will be shared on reasonable request to the corresponding author.




\bibliographystyle{mnras}
\bibliography{bibliography} 

%
%

\appendix

\section{Convergence study}
\begin{figure}
\includegraphics[width=\columnwidth]{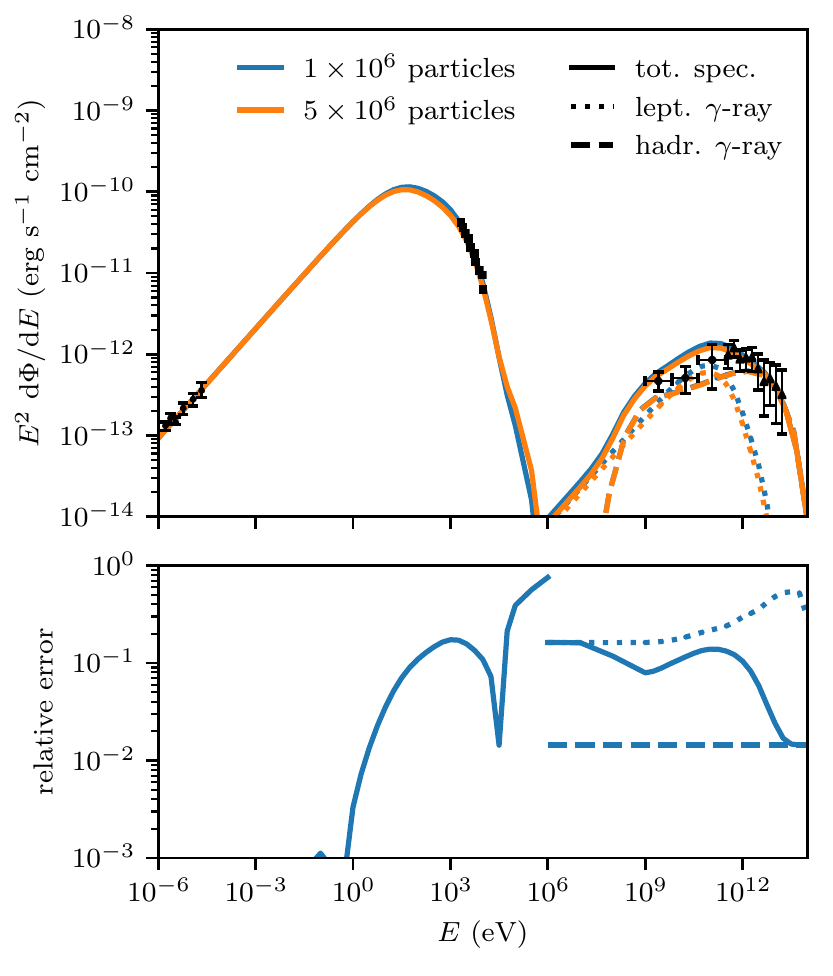}
\caption{Multi frequency spectra (top) and relative error (bottom) for a simulation at
  low resolution with \num{1e6} cells (blue) and at high resolution
  with \num{5e6} cells (orange). The spectrum is calculated with our best-fit
  parameters: spectral indices $\alpha_\mathrm{e}=2.1$ for electrons and
  $\alpha_\mathrm{p}=1.9$ for protons, gas density of
  $n=\SI{0.12}{\per\cubic \centi\meter}$, distance of $D=\SI{1660}{pc}$,
  equivalent magnetic field of $B=\SI{35}{\micro G}$ (as a result of a turbulent
  dynamo), and Bell amplification by a factor of 20.}  \label{fig:Resolution}
\end{figure}

We briefly discuss the numerical convergence of our simulations. As described in section~\ref{sec:ICs}, we use two setups, that only differ in their number of resolution elements (cells and tracer particles). Figure~\ref{fig:Resolution} shows the multi-frequency spectrum (top panel) that is calculated for the low resolution of \num{1e6} cells (blue lines) and the high resolution of \num{5e6} cells (orange lines). Both spectra are calculated with our best-fit parameters which are spectral indices $\alpha_\mathrm{e}=2.1$ for electrons and $\alpha_\mathrm{p}=1.9$ for protons, gas density of $n=\SI{0.12}{\per\cubic \centi\meter}$, distance of $D=\SI{1660}{pc}$, equivalent magnetic field of $B=\SI{35}{\micro G}$ (as a result of a turbulent dynamo), and Bell amplification by a factor of 20.
The bottom panel of Figure~\ref{fig:Resolution} shows the relative error 
\begin{align}
\delta = \left| 1 - \frac{\mathrm{d}\Phi_\mathrm{low}/\mathrm{d}E}{ \mathrm{d}\Phi_\mathrm{high}/\mathrm{d}E} \right|
\end{align}
of the low to high resolution simulation spectrum. The relative error becomes largest in the cutoff regions of the synchrotron and IC spectra. However, the relative error is below 15 per cent at the X-ray and $\gamma$-ray data points. This is accurate enough to enable our parameter space study presented in Section~\ref{sec:parameters} at a feasible computational costs.


\bsp	
\label{lastpage}
\end{document}